\documentclass[10pt,journal]{IEEEtran} 
\usepackage{cite}
\usepackage{threeparttable}
\usepackage{graphicx}
\usepackage{picinpar}
\usepackage{stfloats}
\usepackage{setspace}
\usepackage[all]{xy}
\usepackage{float}
\usepackage[cmex10]{amsmath}
\usepackage{threeparttable}
\usepackage{array}
\usepackage{mdwmath}
\usepackage{mdwtab}
\usepackage{eqparbox}
\usepackage{algorithmic}
\usepackage{algorithm}
\usepackage{exscale}
\usepackage{relsize}
\usepackage{supertabular}
\usepackage{subfigure}
\usepackage{threeparttable}
\usepackage{caption}
\usepackage{booktabs}
\usepackage{multirow}
\usepackage{etoolbox}
\usepackage{makecell}
\usepackage{amssymb}
\usepackage{mathrsfs}
\usepackage{color}
\definecolor{purple}{RGB}{160,32,240}
\definecolor{darkred}{RGB}{255,0,255}
\usepackage{stfloats}
\usepackage{bm}

\usepackage{hyperref}
\usepackage{url}

\usepackage[justification=centering]{caption}

\usepackage{amsthm}

\newtheorem{theorem}{Theorem}
\newtheorem{prop}{Proposition}
 
\newtheorem{corollary}{Corollary}
\newtheorem{remark}{Remark}
\newtheorem{assumption}{Assumption}

\usepackage{smile}

\usepackage{flushend}
\begin{document}

\newtheorem{corol}{Corollary}
\newcommand{\e}{\begin{equation}}
\newcommand{\ee}{\end{equation}}
\newcommand{\eqn}{\begin{eqnarray}}
\newcommand{\eeqn}{\end{eqnarray}}

\renewcommand{\algorithmicrequire}{ \textbf{Input:}}
\renewcommand{\algorithmicensure}{ \textbf{Output:}}

\title{Massive Digital Over-the-Air Computation for Communication-Efficient Federated Edge Learning}

\author{Li Qiao,~\IEEEmembership{Graduate Student Member,~IEEE}, Zhen Gao,~\IEEEmembership{Member,~IEEE}, \\
Mahdi Boloursaz Mashhadi,~\IEEEmembership{Senior Member,~IEEE}, and 
Deniz Gündüz,~\IEEEmembership{Fellow,~IEEE}
 \vspace{-6mm}

\thanks{This work was supported in part by the National Natural Science Foundation of China (NSFC) under Grant U2233216 and Grant 62071044, in part by the Shandong Province Natural Science Foundation under Grant ZR2022YQ62, and in part by the Beijing Nova Program. The work of L. Qiao was supported by the China Scholarship Council, and was carried out while he was a visiting student at Imperial College London. This work received funding from the UKRI for the projects AI-R (ERC Consolidator Grant, EP/X030806/1) and SONATA (EPSRC-EP/W035960/1), and the SNS JU project 6G-GOALS under the EU’s Horizon program (Grant Agreement No. 101139232.) Part of the results were presented at the 2023 IEEE International Symposium on Information Theory (ISIT) \cite{ISIT}. We release our code at {\it https://github.com/liqiao19/MD-AirComp} and {\it https://gaozhen16.github.io/} to facilitate further research and reproducibility. For the purpose of open access, the authors have applied a Creative Commons Attribution (CCBY) license to any Author Accepted Manuscript version arising from this submission. (Corresponding author: Zhen Gao)}

\thanks{L. Qiao is with the School of Information
and Electronics, Beijing Institute of Technology, Beijing 100081, China, also with 5GIC \& 6GIC, the Institute for Communication Systems (ICS), University of Surrey, GU2 7XH Guildford, U.K. (e-mails: qiaoli@bit.edu.cn).}
\thanks{Z. Gao is with Beijing Institute of Technology (Zhuhai), Zhuhai 519088, China, also with the MIIT Key Laboratory of Complex-Field Intelligent Sensing, Beijing Institute of Technology, Beijing 100081, China, also with the Yangtze Delta Region Academy of Beijing Institute of Technology, Beijing Institute of Technology, Jiaxing 314000, China, also with the Advanced Technology Research Institute, Beijing Institute of Technology, Jinan 250307, China. (e-mails: gaozhen16@bit.edu.cn).}
\thanks{M. Boloursaz Mashhadi is with 5GIC \& 6GIC, Institute for Communication Systems (ICS), University of Surrey, GU2 7XH Guildford, United Kingdom (email: {m.boloursazmashhadi}@surrey.ac.uk).}
\thanks{D. Gündüz is with the Department of Electrical and Electronic Engineering, Imperial College London, London SW7 2AZ, U.K. (email: d.gunduz@imperial.ac.uk).}
}

\maketitle

\vspace{-5mm}
\begin{abstract}
Over-the-air computation (AirComp) is a promising technology converging communication and computation over wireless networks, which can be particularly effective in model training, inference, and more emerging edge intelligence applications. AirComp relies on uncoded transmission of individual signals, which are added naturally over the multiple access channel thanks to the superposition property of the wireless medium. Despite significantly improved communication efficiency, how to accommodate AirComp in the existing and future digital communication networks that are based on discrete modulation schemes remains a challenge. This paper proposes a massive digital AirComp (MD-AirComp) scheme that leverages an unsourced massive access protocol to enhance compatibility with both current and next-generation wireless networks. MD-AirComp utilizes vector quantization to reduce the uplink communication overhead, and employs shared quantization and modulation codebooks. At the receiver, we propose a {near-optimal} approximate message passing-based algorithm to compute the model aggregation results from the superposed sequences, which relies on estimating the number of devices transmitting each code sequence, rather than trying to decode the messages of individual transmitters. We apply MD-AirComp to federated edge learning (FEEL), and show that it significantly accelerates FEEL convergence compared to state-of-the-art while using the same amount of communication resources. 
\end{abstract}
\vspace{-2mm}

\IEEEpeerreviewmaketitle
\vspace{-4mm}
\section{Introduction}

With the advancement of the artificial intelligence of things (AIoT), a growing volume of data is being generated by a massive number of Internet-of-things (IoT) devices, which fuels the ongoing revolution in machine learning (ML) and artificial intelligence (AI) \cite{JSAC-Editor, AIoT,Transformer}. Traditionally, data collected by IoT devices is offloaded to the cloud or data centers to train massively large models \cite{FL-IoT}. However, due to concerns over data privacy and increasing computational capabilities of edge devices, AIoT in wireless networks is transitioning away from centralized learning towards distributed learning frameworks \cite{JSAC_OPT,AI-mag, FL-IoT, AIoT-Mahdi,Mag-Deniz}. One of these emerging frameworks is federated edge learning (FEEL), where a central server coordinates multiple AIoT devices in its coverage area to train an ML model using their respective local datasets and computing resources \cite{Deniz2}. As a result, FEEL requires solving the {parallel and distributed optimization problem} while considering the limited shared wireless resources and interference between devices. Model aggregation process in FEEL involves repeated uplink (UL) transmission of high-dimensional local gradients or model updates from tens to hundreds of devices, which causes a heavy burden on the multiple access networks \cite{G_Zhu}. 

\vspace{-3.6mm}
\subsection{Related Work}
To address this problem, the concept of over-the-air computation (AirComp) has been introduced as a means of achieving communication-efficient model aggregation \cite{Deniz2, G_Zhu, Deniz}. In the context of model aggregation, the objective is to compute the average of local model updates, rather than recovering each update individually. In analog AirComp, the fundamental principle involves embracing inter-user interference over the multiple-access channel (MAC) rather than mitigating it. More specifically, every element of the local model update at a device goes through pre-equalization using the inverse of the UL channel gain from the device to the BS, which acts as the parameter server. Subsequently, the pre-equalized elements are modulated onto the transmit waveform's amplitude. Through simultaneous transmission from multiple devices, the receiver can directly obtain the sum of the local model updates from multiple devices by leveraging the superposed waveform over the MAC. To further reduce the communication overhead, by exploiting sparsification and error accumulation, the dimension of the local model updates are reduced before analog transmission in \cite{Deniz}. After analog AirComp, the receiver reconstructs the average of local model updates by using the approximate message passing (AMP) algorithm. 
A higher FEEL accuracy is achieved with limited communication resources in \cite{Yuxuan}, by exploiting time-correlated sparsification on the local model updates. Multi-input multi-output (MIMO)-based AirComp is studied in \cite{YuanmingMIMO}. Random projection technique proposed in \cite{Deniz} is extended to MIMO channel in \cite{CS_FL}. Note that all the participating devices are synchronized in time in \cite{Deniz, Deniz2, G_Zhu, Yuxuan, YuanmingMIMO, CS_FL}. If the synchronization is imperfect, \cite{Yulin2} and \cite{Yulin} considered misaligned analog AirComp to solve the problem. Dynamic scheduling of devices is proposed in \cite{Yuxuan2} to {optimize} the energy efficiency of AirComp in FEEL. {We refer the readers to \cite{ChenMZ} and \cite{surveyAirComp} for a comprehensive overview of FEEL and AirComp. We should emphasize that AirComp is a general technology that turns wireless medium into a joint communication and computation platform. Its benefits can also be applied to other parallel and distributed optimization applications \cite{TradeoffDO}. For instance, AirComp has recently been used for efficient inference at the edge \cite{Selim, KaibinInfer}.}

Despite the significant potential benefits of AirComp, most of the existing wireless networks adopt digital communication protocols (e.g., 3rd Generation Partnership Project standards) as well as hardware \cite{3GPP}. In other words, both existing and next-generation wireless networks may not have the flexibility to support any arbitrary modulation scheme, which is a fundamental requirement for the analog AirComp methods described earlier. As for digital communication, quantization, i.e., the process of mapping continuous values to a finite set of discrete values, is necessary. Quantization of model parameters in federated learning (FL) has been studied extensively, considering rate-limited error-free communication links from devices to the parameter server \cite{FedPAQ, UVeQFed,AdapQuanFL,VecFL,FedDQ,FLqChang}. {Specifically, the authors of \cite{FedPAQ} analyzed the convergence of FL with stochastic uniform quantization. To optimize the convergence speed of FL with quantization, adaptive quantization levels were investigated in \cite{AdapQuanFL} and \cite{FedDQ}. In addition, joint optimization of bandwidth and quantization levels was considered in \cite{FLqChang}. To further reduce the UL communication overhead, vector quantization (VQ) have shown a significant improvement on the communication efficiency in \cite{UVeQFed} and \cite{VecFL}. Different from the orthogonal UL transmission considered in \cite{FedPAQ, UVeQFed,AdapQuanFL,VecFL,FedDQ,FLqChang}, the authors of \cite{Kaibin} proposed one-bit digital aggregation (OBDA) scheme for FEEL, which combines the ideas of AirComp and sign stochastic gradient descent (signSGD). Specifically, one-bit gradient quantization and digital modulation are adopted at devices, and majority-vote based gradient-decoding via AirComp is employed at the central server. Furthermore, non-coherent detector was proposed in \cite{FSKMV} and \cite{DigFL}, enabling scalar quantization-based digital AirComp without the need of channel state information (CSI). Despite its potential, this approach needs to allocate multiple frequencies to represent the quantization levels, thus raising concerns regarding spectral efficiency. In addition, a series of coding schemes recently proposed in \cite{ChannelComp, SumComp,SumCompFL} enables function computation from superposed digital constellations. However, these emerging digital AirComp schemes, i.e., \cite{Kaibin,FSKMV,DigFL,ChannelComp, SumComp,SumCompFL}, only enables scalar quantization, and it is not clear how/if these approaches can be extended to VQ to further reduce the communication overhead. Hence, this necessities a novel communication-efficient digital AirComp scheme that is suitable for any scalar or vector quantization scheme and can be deployed in existing digital communication systems.}

To fill in this gap, we propose a digital scheme that jointly enables random access and computation among massive AIoT devices. We first revisit grant-free random access (GFRA) protocols, which have been recently studied to enable massive IoT access with low signaling overhead and latency \cite{Malong}. Since devices can directly transmit their signals without the permission of the base station (BS), GFRA protocols reduce the signaling overhead. There are mainly two types of GFRA schemes, i.e., sourced massive access (SMA) \cite{Liuliang,Yafeng2, Mei, Yafeng, Qiao2, zhen, zhen2,ISAC} and unsourced massive access (UMA) \cite{WuJSAC, Yury, Shao_UMA, Durisi}. Specifically, the SMA schemes proposed in \cite{Liuliang,Mei,zhen} have demonstrated their superiority in both access latency and detection performance with the help of the elegant AMP algorithm framework. Furthermore, authors in \cite{Qiao2} and \cite{zhen2} extend the SMA framework to non-terrestrial networks, which efficiently improve the transmission efficiency and reduce the detection error at the same time. Both SMA and UMA schemes adopt non-orthogonal random access codebook for reducing the communication overhead. In SMA, each device has its unique non-orthogonal preamble indicating its identity. The BS performs active device detection and channel estimation, followed by data decoding \cite{Liuliang,Yafeng2, Mei, Yafeng, Qiao2, zhen, zhen2,ISAC}. In UMA, instead, all the devices adopt the same non-orthogonal codebook. Transmitted bits are modulated by the index of the UMA codewords and decoded at the BS. The goal of the BS is to decode the list of transmitted messages, but not the 
identities of the active devices \cite{WuJSAC, Shao_UMA,Yury,Durisi}. We remark that the identities of active devices are not necessary for model aggregation in FEEL either. Inspired by UMA, our target is to redesign the modulation and decoding modules of conventional UMA for efficient AirComp across a large number of devices.

\vspace{-4mm}
\subsection{Contributions}
To improve the communication efficiency of FEEL, we propose a massive digital AirComp (MD-AirComp) scheme that can be deployed in existing and next-generation digital communication systems. Specifically, consider a scenario with multiple devices, each with its own signal vector. The goal of the receiver is to recover the average of these vectors. In MD-AirComp, individual signal vectors are quantized using the common VQ codebook. Then, the active devices select their transmit sequences from the common modulation codebook based on their quantization indices. Note that there is a one-to-one mapping between the quantization and modulation codebooks. After pre-equalization, the non-orthogonal sequences from active devices are transmitted and overlap at the BS. Then, an AMP-based digital aggregation (AMP-DA) algorithm is proposed at the BS to estimate the number of devices employing each modulation codeword. We remark that in most FEEL and edge inference applications, active devices are likely to transmit the same message, especially when their datasets/models are statistically similar \cite{CS_FL,Selim}. Our main contributions are summarized as follows:
\begin{itemize} 
\item[$\bullet$] {\color{black}\bf Communication-efficient MD-AirComp method:} By exploiting the superposition property of MAC and VQ, an MD-AirComp scheme is proposed for communication-efficient FEEL across digital communication systems. Since the receiver is interested in the sum of the signal vectors, the MD-AirComp scheme is user-collision friendly. {\color{black}In addition, it can be deployed with any scalar or vector quantization scheme to exploit the distribution of the signal vectors for better training accuracy. Since our proposed MD-AirComp scheme is a general lossy computing scheme over a MAC, it can also be applied to other edge computing tasks, such as collaborative inference \cite{Selim}.}

\item[$\bullet$] {\bf Proposed AMP-DA algorithm for signal aggregation:} {\color{black}Due to the sporadic activity of AIoT devices and the potentially high-resolution VQ in FEEL, the signal to be detected is inherently sparse. Moreover, due to similarities in local dataset, some devices are likely to transmit the same message, and user-collisions will make the signal even more sparse. In addition, non-zero elements in the signal are integers representing the numbers of devices selecting a specified codeword. By modeling these properties in the prior distribution, the proposed AMP-DA algorithm realizes efficient model aggregation. Furthermore, the AMP-DA algorithm uses majority voting to estimate the number of active devices and exploits the multiple observations from multi-antenna BS, to further enhance the detection accuracy.}

\item[$\bullet$] {\bf Convergence analysis of MD-AirComp-based FEEL:} We analyze the convergence rate of FEEL when the proposed MD-AirComp scheme is employed. Specifically, we analyze the influence of errors, originating from VQ and wireless communication, on the convergence rate. The wireless communication error relates to the detection accuracy of the AMP-DA algorithm. Our findings indicate that the MD-AirComp-based FEEL achieves a convergence rate of $\cO(\frac{1}{\sqrt{T}})$, where $T$ is the number of global training rounds.

\end{itemize}

\textit {Notation}: Boldface lower and upper-case symbols denote column vectors and matrices, respectively. For a matrix ${\bf A}$, ${\bf A}^T$, ${\bf A}^*$, ${\bf A}^H$, ${\left\| {\bf{A}} \right\|_F}$, $[{\bf{A}}]_{m,n}$ denote the transpose, conjugate, Hermitian transpose, Frobenius norm, and the $m$-th row and $n$-th column element of ${\bf{A}}$, respectively. $[{\bf{A}}]_{:,n}$ denotes the $n$-th column of ${\bf A}$. For a vector ${\bf x}$, $\| {\bf x} \|_p$ and $[{\bf x}]_{m}$ denote the ${l_p}$ norm and $m$-th element of ${\bf x}$, respectively. $|\Gamma|$ denotes the cardinality of the ordered set $\Gamma$. $\lfloor \cdot \rfloor$ ($\lceil \cdot \rceil$) rounds each element to the nearest integer smaller (larger) than or equal to that element. The marginal distribution $p\left([{\bf x}]_m\right)$ is denoted as $p\left([{\bf x}]_m\right)={\int}_{\backslash [{\bf x}]_m} p\left({\bf x}\right)$. $\mathcal{CN}(x; \mu, \nu)$ or $x\sim\mathcal{CN}(\mu, \nu)$ denotes the complex Gaussian distribution of random variable $x$ with mean $\mu$ and variance $\nu$. $[K]$ denotes the set $\{1,2,...,K\}$. ${\bf 0}_{m\times 1}$ denotes the all-zero vector with $m$ rows.
\begin{figure}[t]
     \centering
     \includegraphics[width = 0.67\columnwidth,keepaspectratio]{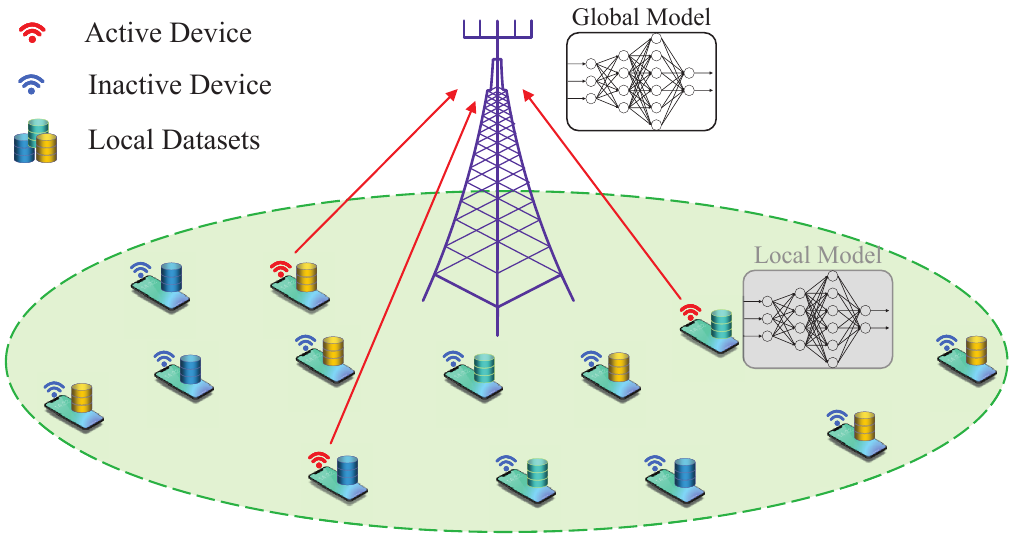}
     \captionsetup{font={footnotesize, color = {black}}, singlelinecheck = off, justification = raggedright,name={Fig.},labelsep=period}
     \caption{Illustration of the FEEL scenario in cellular networks.}
     \label{fig1}
     \vspace{-7mm}
\end{figure}

\vspace{-2.7mm}
\section{System Model}
As shown in Fig. \ref{fig1}, we consider $K$ single-antenna edge devices served by an $M$-antenna BS in a cellular system. Each device $k$, $k\in[K]$, has its own local dataset $\mathcal{B}_k$, which consists of labeled data samples. For convenience, we assume uniform sizes for local datasets, that is, $|\mathcal{B}_k|=B$, $\forall k$. {In FEEL, $K$ devices are coordinated by the the BS to solve the following distributed optimization problem in parallel
\vspace{-2mm}
\begin{align}\label{FEELproblem}
\text{min}_{{\bf w}\in\mathbb{R}^W} f(\wb)=\frac{1}{K}\sum\nolimits_{k=1}^KF_k(\wb),
\end{align}
where ${\bf w}\in\mathbb{R}^W$ is the parameter vector of a common neural network model and $F_k(\wb)$ is the local loss function.}

{We follow the seminal federated averaging (FedAvg) \cite{FedAvg} algorithm\footnote{Although we consider FedAvg in this paper, an extension
of the results to other federated optimization algorithms \cite{Fedboost, FedProx, FedNova} is straightforward.}.} Specifically, in the ($t-1$)-th global training round of FEEL, $\forall t\in[T]$, the BS broadcasts the current global model ${\bf w}^{t-1}$ to the devices. Based on ${\bf w}^{t-1}$ and its local dataset $\mathcal{B}_k$, the $k$-th device performs $T_l$ local iterations of stochastic gradient
descent (SGD) to obtain ${\bf w}_k^{t}$, which is expressed as 
\begin{align}\label{LossGra}
    {\bf w}_{k, {t_l}}^{t}\!=\!{\bf w}_{k, {t_l-1}}^{t}\!-\!\dfrac{\eta_l}{|\mathcal{B}_k|}\!\sum_{{\bm \varepsilon}\in\mathcal{B}_k} \!\nabla F_k({\bf w}_{k, {t_l-1}}^{t},{\bm \varepsilon}),
\end{align}
where ${t_l}\in[T_l]$, ${\bf w}_{k, 0}^{t}={\bf w}^{t-1}$, ${\bf w}_k^{t}={\bf w}_{k, T_l}^{t}$, $\nabla$ and $\eta_l$ denote the gradient operator and the local learning rate, respectively, $F_k({\bf w}_{k, {t_l}-1}^t,{\bm \varepsilon})$ denotes the local loss of model ${\bf w}_{k, {t_l}-1}^t$ on training sample ${\bm \varepsilon}$ with respect to its true label.  

According to (\ref{LossGra}), the local model update can be obtained as $\Delta_k^t = {\bf w}_{k}^t - {\bf w}^{t-1}$, $\forall k,t$. Ideally, the local model updates will be transmitted to the BS for global model update. In practice, to reduce the communication cost of $\Delta_k^t$, compression\footnote{Note that the typically used quantization and sparsification are two kinds of compression methods.} and error accumulation are commonly adopted in FEEL \cite{Deniz}. In particular, 
we denote $\overline{\bf s}_k^t = \Delta_k^t+{\bf e}_k^t$, $\forall k,t$, where the local model update $\Delta_k^t$ is compensated with the accumulated error vector ${\bf e}_k^t$. In addition, ${\bf e}_k^t$ is defined as ${\bf e}_k^{t}=\overline{\bf s}_k^{t-1}-\mathcal{C}(\overline{\bf s}_k^{t-1})$, where $\mathcal{C}(\cdot)$ denotes compression on its argument and the output is the element-wise compressed values. It can be seen that the accumulated
error vector keeps track of the values that are not updated to the
BS due to compression, helping to accelerate the training \cite{ErrAccum}. Hence, the compressed local model update ${\bf s}_k^t=\mathcal{C}(\overline{\bf s}_k^t)$ is transmitted to the BS rather than $\Delta_k^t$, $\forall k,t$, to reduce the communication cost while maintaining good training accuracy.

Furthermore, it is commonly adopted that only $K_a$ ($K_a\ll K$) devices are simultaneously active in AIoT networks {\cite{Deniz, FLqChang,surveyAirComp}}. We assume the set of active devices $\mathcal{S}_a$, where $|\mathcal{S}_a|=K_a$, send their compressed local model updates ${\bf s}_k^t$, $k\in\mathcal{S}_a$, to the BS in each round. If ${\bf s}_k^t$ can be perfectly obtained at the BS, the global model update ${\bf s}^t\in\mathbb{R}^W$ can be calculated as
\begin{align}\label{GlobalGra}
    {\bf s}^t=\dfrac{1}{K_a}\sum\nolimits_{k\in\mathcal{S}_a} {\bf s}_k^t.
\end{align}
Then, the global model can be updated as
\begin{align}\label{GlobalWe}
    {\bf w}^{t+1}={\bf w}^{t} + \eta{\bf s}^t,
\end{align}
where $\eta$ is the global learning rate. Finally, the updated parameter vector ${\bf w}^{t+1}\in\mathbb{R}^W$ is sent back to the devices, and steps (\ref{LossGra}), (\ref{GlobalGra}), and (\ref{GlobalWe}) are iterated until a convergence condition is met.

It is clear from (\ref{GlobalGra}) that only the sum of the local model updates, rather than the individual values, is needed at the BS. The BS does not need to identify the active devices either. However, due to the large dimension of the parameter vector and the large number of devices that can potentially participate in the training, there is a communication bottleneck in FEEL. {\color{black}Thereby, we propose a communication-efficient MD-AirComp scheme as explained in Section III.}

\begin{remark}
\label{UEspar}
One reasons to have $K_a\ll K$ in FEEL could be due to heterogeneous computation speeds across edge devices \cite{Deniz, FLqChang,surveyAirComp, FedNova}. For a given time slot, only the devices that have finished their computation task will transmit. Alternatively devices can join or leave the learning process at different frames based on their energy or data availability as considered in typical IoT scenarios \cite{JSAC-Editor}.
\end{remark}

\vspace{-2mm}
\section{The Proposed MD-AirComp Scheme}
\begin{figure}[t] 
\vspace{-2mm} 
\centering  
\includegraphics[width = 0.9\columnwidth]{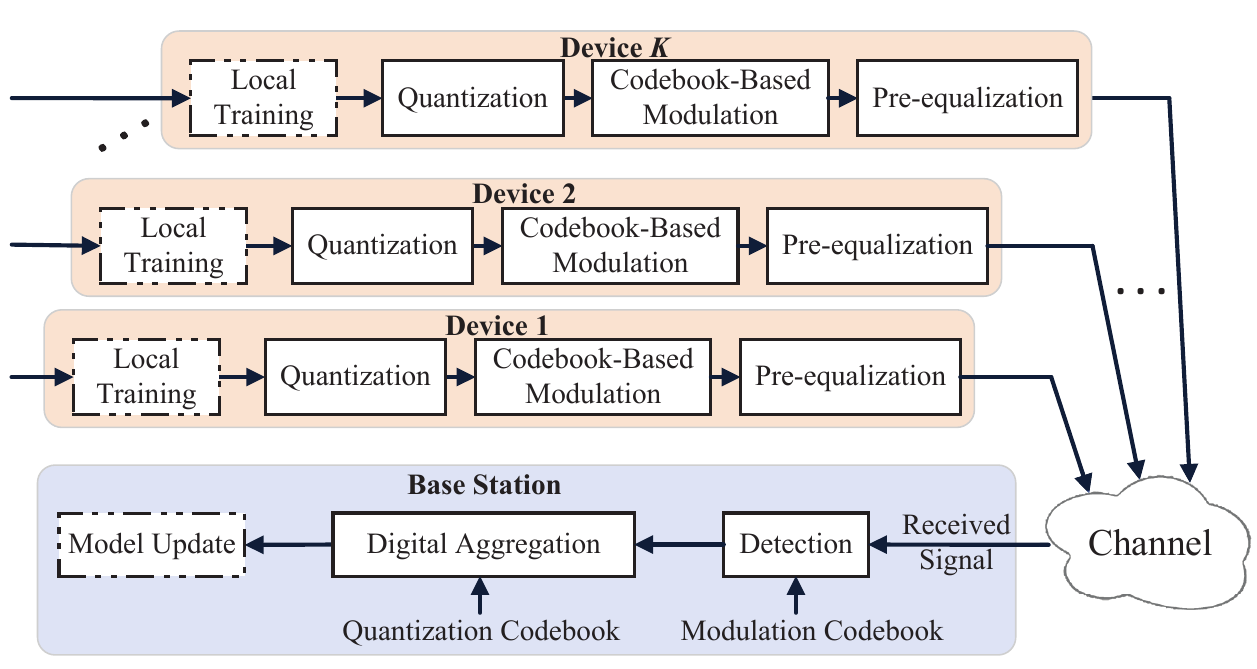} 
\captionsetup{font={footnotesize}, singlelinecheck = off, justification = justified,name={Fig.},labelsep=period}
\caption{The schematic diagram of the proposed MD-AirComp scheme, where the modules with dotted boxes are specific to FEEL.} 
\vspace{-7mm}
\label{fig2}
\end{figure}

{\color{black}As shown in Fig. \ref{fig2}, the proposed MD-AirComp scheme consists of several modules, i.e., a quantization module, a codebook-based modulation module, and a pre-equalization module at the transmitter, a detection module and a digital aggregation module at the receiver. In addition, we use dotted boxes to describe the local training module and the model update module that are specific to FEEL. In the following, we will describe each of these modules in detail. For simplicity, we omit the index $t$ unless otherwise specified. We reemphasize that MD-AirComp is a general approach for lossy computation over a MAC. In this paper, we motivate it through its application to FEEL.}

\vspace{-2mm}
\subsection{Transmitter Design}
At the transmitter, each active device first quantizes $\overline{\bf s}_k\in\mathbb{R}^W$, and then modulates the quantized value to a common modulation codebook. All active devices transmit their modulation codewords simultaneously. Next, we explain each component in detail.

\subsubsection{Quantization Design}
We will employ a single quantization codebook to be used by all the devices, denoted by ${\bf U}\in\mathbb{R}^{Q\times N}$, where $N=2^J$ corresponds to $J$-bit quantization with $N$ quantization codewords, $Q\geq 1$ ($Q\in\mathbb{N}$) denotes the length of each quantization codeword, i.e., ${\bf u}_n\in\mathbb{R}^{Q}$, $\forall n\in[N]$. Note that $Q=1$ and $Q>1$ indicate scalar quantization and VQ, respectively.

The vector $\overline{\bf s}_k\in\mathbb{R}^W$ at device $k$ is divided into $D=W/Q$ blocks. For simplicity, here we assume that $W$ can be divided by $Q$. Each block is then quantized independently using codebook ${\bf U}$. Let $\overline{\bf s}_k\in\mathbb{R}^W$ be quantized into ${\bf b}_k\in[N]^D$, where the $d$-th element of ${\bf b}_k$, $[{\bf b}_k]_d$, denotes the quantization index of the corresponding block. In particular, $[{\bf b}_k]_d$ is identified by finding the quantization codeword with the minimum Euclidean distance, which can be expressed as
\begin{align}\label{Quant}
    \left[{\bf b}_k\right]_d= \arg\min_{i\in[N]} \left\| \left[{\bf U}\right]_{:,i} - \left[\overline{\bf s}_k\right]_{(d-1)Q+1:dQ}\right\|_2.
\end{align}
To determine the quantization codebook, we can use any VQ method. From (\ref{Quant}), it can be seen that the size of each local model update can be reduced by using VQ, thereby decreasing the communication overhead in the UL by a factor of $Q$.

Furthermore, we introduce a one-hot vector ${\bf x}_k^d\in\{0,1\}^N$, namely the selection vector, where $\|{\bf x}_k^d\|_0=1$ and $\left[{\bf x}_k^d\right]_{\left[{\bf b}_k\right]_d}=1$. Hence, we can denote the quantized version of $\overline{\bf s}_k$ as
\begin{align}\label{ErrorAccu}
    {\bf s}_k = \mathcal{C}(\overline{\bf s}_k) = \left({\bf I}_D\otimes{\bf U}\right) \left[ ({\bf x}_k^1)^T,...,({\bf x}_k^D)^T  \right]^T,
\end{align}
where $\otimes$ and ${\bf I}_D$ denote the Kronecker product and the identity matrix of size $D$, respectively.

\subsubsection{Codebook-Based Modulation}
We consider the same random access codebook shared by all the devices, denoted by ${\bf P}\in\mathbb{C}^{L\times N}$, where each column of ${\bf P}$ is a codeword of length $L$, and there are $N$ codewords in total. In the proposed scheme, there is a one-to-one mapping between ${\bf P}$ and ${\bf U}$. To be specific, if the device acquires the quantized value $[{\bf U}]_{:,n}$, then the modulation codeword (or sequence) $[{\bf P}]_{:,n}$ with the same index $n$ is transmitted. The transmission process is similar to the preamble transmission of random access procedure in 5G communication systems \cite{3GPP}. Hence, the proposed codebook-based modulation is more suitable for the current wireless networks compared to analog modulation typically employed in AirComp. 

\subsubsection{UL Massive Access Model}
We consider a time division duplex system, where the UL and downlink (DL) channels are commonly considered the same due to channel reciprocity \cite{Mei,Durisi}. Before each UL communication round of FEEL, the BS broadcasts a pilot signal and each active device estimates its DL channel based on this pilot signal. Then, the active devices start their grant-free UL transmissions with pre-equalization in a synchronized manner \cite{Mei}. For the $d$-th element of ${\bf b}_k$, $\forall k, d$, the received signal ${\bf Y}_{d}\in\mathbb{C}^{L\times M}$ at the BS can be expressed as
\begin{align}\label{TranMIMO}
    {\bf Y}_{d}=\sum\limits_{k\in\mathcal{S}_a}\dfrac{1}{[{\bf h}_k]_1}{\bf P}{\bf x}_k^d \left({\bf h}_k\right)^T + {\bf Z}_{d}={\bf P}{\bf X}_{d} + {\bf Z}_{d},
\end{align}
where the elements of the UL channel vector ${\bf h}_k\in\mathbb{C}^{M}$ are assumed to follow a complex Gaussian distribution with zero mean and unit variance, while channel inversion is considered for the first antenna of the BS\footnote{Since each device has a single antenna, it is only possible to apply channel inversion for one of the antennas at the BS. For simplicity, we consider channel inversion for the first antenna of the BS. Also, we assume the channel reciprocity and perfect DL channel estimation at the devices \cite{Mei}. Imperfect channel estimation will be discussed in Section VI-E.}. ${\bf X}_{d}=\sum\nolimits_{k\in\mathcal{S}_a}\frac{1}{[{\bf h}_k]_1}{\bf x}_k^d ({\bf h}_k)^T\in\mathbb{C}^{N\times M}$ is the equivalent transmit signal, and elements of the noise vector ${\bf Z}_{d}\in\mathbb{C}^{L\times M}$ obeys the independent and identically distributed (i.i.d.) complex Gaussian distribution with zero mean and variance $\sigma_n^2$. We define the signal-to-noise ratio (SNR) $\text{SNR}\triangleq 10\text{log}_{10}\frac{\|{\bf P}{\bf X}_{d}\|_F^2}{L M \sigma_n^2}$. Hence, the first column of ${\bf X}_{d}$ obeys the following properties,
\begin{align}\label{SpaX}
    \left\|\left[ {\bf X}_{d}\right]_{:,1}\right\|_1=K_a,~\left\|\left[ {\bf X}_{d}\right]_{:,1}\right\|_0\leq K_a,~\left[ {\bf X}_{d}\right]_{n,1}\in \Omega, \forall n,
\end{align}
where $\Omega=\{0, [K_a]\}$. It can be seen that the $n$-th column of ${\bf P}$ is transmitted by $[{\bf X}_d]_{n,1}$ active devices, or equivalently the $n$-th quantization codeword appears at $[{\bf X}_d]_{n,1}$ devices, which indicates how to conduct the model aggregation. The non-zero elements of $[{\bf X}_{d}]_{:,m}$, $m\in\{2,3,...,M\}$, obey the ratio distribution of two uncorrelated complex Gaussian variables as discussed in \cite{RatioDist}. Furthermore, due to the multiple observations at multiple antennas at the BS, ${\bf X}_{d}$ has row sparsity \cite{Liuliang}, which can be expressed as
\begin{align}\label{SparseMIMO}
    \text{supp}\left\{\left[ {\bf X}_{d}\right]_{:,1}\right\}=\text{supp}\left\{\left[ {\bf X}_{d}\right]_{:,2}\right\}=\cdots=\text{supp}\left\{\left[ {\bf X}_{d}\right]_{:,M}\right\},
\end{align}
where $\text{supp}\left\{\cdot\right\}$ denotes the index of non-zero elements (rows) of a vector (matrix). Both (\ref{SpaX}) and (\ref{SparseMIMO}) will be exploited to reconstruct $\left[ {\bf X}_{d}\right]_{:,1}$, $\forall d$, in Section IV.

\begin{remark}
\label{CSIconditionBad}
As shown in (\ref{TranMIMO}), the noise power can be defined by the SNR and the average power of the received signal, which is commonly used in massive access system models \cite{Malong}. If the wireless channel $[{\bf h}_k^{\rm UL}]_1$ is in deep fading, the power of $[{\bf h}_k^{\rm UL}]_1$ can be extremely small. In this case, pre-equalization might increase the received signal power at the rest of the antennas. As a result, SNR at the first antenna can be too small to conduct proper signal detection. In addition, if the channel power is too small, the transmit power needed for pre-equalization might be too large. Hence, we set a threshold $\epsilon_h$ on the channel condition, such that, if $|[{\bf h}_{k}]_1|<\epsilon_h$, the device does not participate in this training round.
\end{remark}

\subsection{Receiver Design}
According to (\ref{GlobalGra}), the receiver's target in MD-AirComp is to compute the aggregated global model update ${\bf s}\in\mathbb{R}^{W}$, which can be expressed as 
\begin{align}\label{computingTask}
    {\bf s} &= \dfrac{1}{{K}_a}\left({\bf I}_D\otimes{\bf U}\right) \left[ \left(\sum\limits_{k\in\mathcal{S}_a}{\bf x}_k^1\right)^T,...,\left(\sum\limits_{k\in\mathcal{S}_a}{\bf x}_k^D\right)^T   \right]^T \nonumber \\ 
    & = \dfrac{1}{{K}_a}\left({\bf I}_D\otimes{\bf U}\right) \left[ \left( 
 \left[ {\bf X}_{1}\right]_{:,1}\right)^T,...,\left(  \left[ {\bf X}_{D}\right]_{:,1} \right)^T   \right]^T.
\end{align}
It can be seen that the main differences between (\ref{computingTask}) and the decoding problem of UMA are as follows
\begin{itemize}
    \item {\bf The proposed MD-AirComp scheme is collision friendly:} In particular, the superposed equivalent signal vector $[ {\bf X}_{d}]_{:,1}$ is used for the computing task (\ref{computingTask}) as a whole. Hence, it is unnecessary to decode each device's data from $[ {\bf X}_{d}]_{:,1}$, $\forall d$. As a result, it is unnecessary to avoid collisions of multiple devices.  
    \item {\bf Detection algorithms for UMA do not apply to MD-AirComp:} In UMA, since data is embedded in the index of the transmitted codeword, detection algorithms focus on estimating the support of ${\bf X}_d$ rather than its values, $\forall d$. Furthermore, the number of active users $K_a$ is trivial in UMA, but its detection is crucial for (\ref{computingTask}). In addition, due to pre-equalization in MD-AirComp, the received signal distributions on the first antenna and the remaining antennas are different. 
\end{itemize}

\subsubsection{CS-Based Detection}
To achieve higher quantization accuracy, especially for VQ, the number of quantization bits $J$ can be set very large, e.g., more than 8 bits. In this case, $N=2^J$ can be far greater than $K_a$ (i.e., $K_a \ll N$), which motivates us to employ non-orthogonal codebook ${\bf P}$ ($L<N$) and CS-based detection algorithms to reduce the communication overhead. Thanks to the sparsity of the equivalent signal $[ {\bf X}_{d}]_{:,1}$, i.e., the number of non-zero elements $\|[ {\bf X}_{d}]_{:,1}\|_0\leq K_a \ll N$, we can employ CS-based algorithms to estimate ${\bf x}_d$. Note that the codebook ${\bf P}$ is known at the BS. Our design target is to estimate the elements $[ {\bf X}_{d}]_{:,1}$ as accurate as possible. The details of the proposed CS-based detection algorithm will be given in Section IV.

Furthermore, since the number of non-zero elements $\|[ {\bf X}_{d}]_{:,1}\|_0$ will decrease if more users transmit the same codeword, the interference improves sparsity thereby becomes helpful for the CS-based detection. Hence, unlike the traditional UMA \cite{WuJSAC}, user collision avoidance is unnecessary.

\begin{algorithm}[t]
\algsetup{linenosize=\small} \small
\color{black}
\caption{Proposed MD-AirComp Scheme}\label{Algorithm:1}
\begin{algorithmic}[1] 
\raggedright 
\STATE {\bf Initialization:} Initialize model ${\bf w}^{0}\in\mathbb{R}^W$, accumulated error vector ${\bf e}_k^0={\bf 0}_{W\times1}$, $\forall k\in[K]$, and global training round $t=1$.
\WHILE{$t<T$} 
\STATE{BS broadcasts ${\bf w}^t$ to all the devices.} 
\STATE Local training via (\ref{LossGra}) to obtain $\Delta_k^t$;
\STATE Error accumulation $\overline{\bf s}_k^t = \Delta_k^t+{\bf e}_k^t$;
\STATE Quantization via (\ref{Quant}) and (\ref{ErrorAccu}) to obtain $({\bf x}_k^d)^t$ and ${\bf s}_k^t$;
\label{QAT}
\STATE Refine the accumulated error vector ${\bf e}_k^{t+1}=\Delta_k^t+{\bf e}_k^t-{\bf s}_k^t$;
\STATE Based on the quantization results in line \ref{QAT}, each device selects modulation codewords ${\bf P}({\bf x}_k^d)^t$, $\forall d,k$;
\STATE BS broadcasts pilot signal for synchronization and channel pre-equalization. Active devices transmit, as shown in (\ref{TranMIMO});
\STATE BS estimates overlapped selection vector $[{\bf X}_d]_{:,1}$, $\forall d$, and number of active devices $K_a$;
\STATE Model aggregation via (\ref{computingTask});
\STATE Model update via (\ref{GlobalWeEs}).
\STATE $t=t+1$.
\ENDWHILE
\end{algorithmic}
\end{algorithm}

\subsubsection{Digital Aggregation}
After CS-based detection, we can obtain an estimate of the equivalent transmit signals, denoted by $[ \widehat{\bf X}_{d}]_{:,1}$, $\forall d$. According to (\ref{SpaX}), estimated number of active devices $\widehat{K}_a$ can be obtained by calculating $\|[ \widehat{\bf X}_{d}]_{:,1}\|_1$, $\forall d$, which will be illustrated in Section IV-C in detail. Then, by substituting $\widehat{K}_a$ and $[ \widehat{\bf X}_{d}]_{:,1}$ into (\ref{computingTask}), we obtain the estimated global model update $\widehat{\bf s}$.

\subsubsection{Model Update}
According to (\ref{GlobalWe}) and (\ref{computingTask}), BS updates the weight of the ML model as
\begin{align}\label{GlobalWeEs}
    \widehat{\bf w}^{t+1}=\widehat{\bf w}^{t} + \eta \cdot \widehat{\bf s}^t,
\end{align}
and broadcasts it to all the devices and the next training round starts. Finally, the proposed MD-AirComp based FEEL procedure is summarized in {\bf Algorithm} \ref{Algorithm:1}.

\vspace{-2.5mm}
\section{Proposed CS-Based Digital Aggregation Algorithm}
In this section, we will first introduce the CS-based problem formulation. Furthermore, we propose the AMP-based algorithm for efficient detection and $K_a$ estimation. Finally, we summarize the proposed AMP-DA algorithm and analyse its computational complexity. For simplicity, we omit the subscript $d$ in this section unless otherwise specified.

\vspace{-4.5mm}
\subsection{Problem Formulation}    
According to (\ref{computingTask}), since ${\bf U}$ and ${\bf P}$ are known at the receiver, the target of the BS is to estimate $[{\bf X}]_{:,1}$ and $K_a$ from the observation ${\bf Y}$, as shown in (\ref{TranMIMO}). {Hence, the detection problem at the BS can be formulated as follows}
\begin{equation}
\begin{aligned} \label{P}
&\min_{{\bf X},K_a} \quad \left\| {\bf Y}-{\bf PX}\right\|_F^2\\
&\begin{array}{r@{\quad}r@{}l@{\quad}l}
s.t. \quad (\ref{SpaX}), \quad (\ref{SparseMIMO}), \quad \text{and} \quad K_a \ll N.
\end{array}
\end{aligned}
\end{equation}
{The optimization problem (\ref{P}) minimizes the mean square error (MSE) between ${\bf Y}$ and ${\bf PX}$, which is equivalent to calculate the posterior mean of ${\bf X}$ under the Bayesian framework \cite{SMKay}.} In the following, we denote $[{\bf X}]_{n,m}$ as $x_{n,m}$, where $n\in[N]$ and $m\in[M]$. Furthermore, the posterior mean of $x_{n,m}$, $\forall n,m$, can be expressed as
\begin{align}\label{eq:Pmean}
    \widehat{x}_{n,m}= {\int} x_{n,m} p\left(x_{n,m}\vert [{\bf Y}]_{:,m}\right)dx_{n,m},
\end{align}
where $p(x_{n,m}| [{\bf Y}]_{:,m})$ is the marginal distribution of $p([{\bf X}]_{:,m}| [{\bf Y}]_{:,m})$ and it can be written as
\begin{align}\label{eq:Marginal}
\vspace{-3mm}
    p\left(x_{n,m}| [{\bf Y}]_{:,m}\right)= \mathlarger{\int}_{\backslash x_{n,m}} p\left([{\bf X}]_{:,m}| [{\bf Y}]_{:,m}\right).
\vspace{-3mm}
\end{align}

As discussed in Section III-B, to achieve higher quantization accuracy, more quantization bits are needed for higher dimensional VQ. This can result in a large dimension of $N$, which demands a computationally efficient algorithm to calculate the marginal distribution $p(x_n| {\bf{y}})$. To address this issue, the AMP algorithm can be adopted to obtain the approximate marginal distributions with relatively low complexity, {while achieving near minimum mean square error (MMSE) performance} \cite{Donoho}. 

\vspace{-4mm}
\subsection{Approximate Message Passing (AMP) Update Rules}

According to the AMP algorithm \cite{Donoho}, in the large system regime, the estimation problem (\ref{TranMIMO}) can be approximately decoupled into $N$ scalar problems as
\begin{equation}\label{eq:Decoupling}
\begin{array}{l}
{\bf Y}={\bf P}{\bf{X}} +{\bf{Z}}\; \to \;r_{n,m} = x_{n,m} + z_{n,m}, ~\forall n, m,
\end{array}
\end{equation}
where $z_{n,m}\sim {\cal CN}(0,\varphi_{n,m})$, $r_{n,m}$ is the estimation of $x_{n,m}$. Hence, according to Bayes's theorem, the posterior distribution of $x_{n,m}$ can be expressed as
\begin{align}\label{eq:Bayes}
    p\left(x_{n,m}| [{\bf Y}]_{:,m}\right)\! &\approx \!p\left(x_{n,m}| r_{n,m}\right)\!\nonumber\\
    &= \!\!\dfrac{1}{p\left(r_{n,m}\right)}p\left(r_{n,m}|x_{n,m}\right)p\left(x_{n,m}\right),
\end{align}
where symbol ``$\approx$" is due to the approximation. The likelihood function and the normalization term are, respectively, denoted as
\begin{align}
    p\left(r_{n,m}|x_{n,m} \right)&= {\cal CN}\left(r_{n,m};x_{n,m},\varphi_{n,m}\right), \label{eq:BayesDetail1}\\  
    p\left(r_{n,m} \right)&= {\int}p\left(r_{n,m}|x_{n,m}\right)p\left(x_{n,m}\right)dx_{n,m}.
    \label{eq:BayesDetail2}
\end{align}
During the $i$-th iterations of AMP, $(r_{n,m})^i$ and $(\varphi_{n,m})^i$, $\forall n,m$, are updated at the variable nodes of the factor graph as \cite{AMPmeng,AMP2}
\begin{align}
\label{eq:UpdateSigma}(\varphi_{n,m})^i&=\left({\sum\nolimits_{l=1}^{L}\dfrac{\left|[{\bf P}]_{l,n}\right|^2}{\sigma^2+(V_{l,m})^i}}\right)^{-1},\\
\label{eq:UpdateR} (r_{n,m})^{i}&\!\!=\!\!(\widehat{{x}}_{n,m})^{i}\!\!+\!\!(\varphi_{n,m})^{i}\!\sum\limits_{l=1}^{L}\!\!\!\dfrac{[{\bf P}]_{l,n}^*\left({\left[{\bf Y}\right]_{l,m}\!\!\!-\!\!(Z_{l,m})^{i}}\right)}{\sigma^2+(V_{l,m})^{i}},
\end{align}
where $(V_{l,m})^{i}$ and $(Z_{l,m})^{i}$, $\forall l,m$, are updated at the factor nodes as
\begin{align}
\label{eq:UpdateV} (V_{l,m})^{i}&=\sum\nolimits_{n=1}^N\left|[{\bf P}]_{l,n}\right|^2 (\widehat{v}_{n,m})^{i},\\
\label{eq:UpdateZ} (Z_{l,m})^{i}&=\sum\limits_{n=1}^N [{\bf P}]_{l,n}(\widehat{x}_{n,m})^{i}-(V_{l,m})^{i}\dfrac{\left[{\bf Y}\right]_{l,m}-(Z_{l,m})^{i-1}}{\sigma^2+(V_{l,m})^{i-1}}.
\end{align}
Note that $(\cdot)^i$ denotes its argument in the $i$-th AMP iteration, ${\widehat x}_{n,m}$ and ${\widehat v}_{n,m}$ denote the posterior mean and variance of $x_{n,m}$, respectively. 

Furthermore, we delve into the prior distribution $p(x_{n,m})$ needed for the posterior estimation (\ref{eq:Bayes}). According to (\ref{SpaX}), we model the prior distribution of $x_{n,1}$ and $x_{n,m}$, $m\in\{2,3,...,M\}$, as 
\begin{align}
    \label{eq:prior1}p\left(x_{n,1} \right)\!= \!\left(1\!-\!a_{n,1} \right)\delta(x_{n,1}) \!+ \!\dfrac{a_{n,1}}{K_a}\!\!\sum\limits_{s\in[K_a]}\delta(x_{n,1}-s),
\end{align}
\begin{align}
    \label{eq:prior2}p\left(x_{n,m}\right)\!=\!(1\!-\!a_{n,m})\delta\left( x_{n,m}\right)\!\!+\!a_{n,m}{\cal CN}\left( x_{n,m}; \!{\mu}_0,  \!{\tau}_0 \right),
\end{align}
respectively, where the sparsity indicator $a_{n,m}$ ($a_{n,1}$) equals to zero if $x_{n,m}$ ($x_{n,1}$) equals to zero, otherwise $a_{n,m}$ ($a_{n,1}$) equals to one. Also, in (\ref{eq:prior1}), if $a_{n,1} = 1$, we assume that $x_{n,1}$ can be any element from set $[K_a]$ with equal probability. Note that this assumption is an approximation since the BS cannot obtain the actual distribution of each element $x_{n,1}$, $\forall n\in[N]$. By adopting (\ref{eq:prior1}), the algorithm tends to obtain estimates of $x_{n,1}$ (if $a_{n,1} = 1$) that are integers from $[K_a]$.

Substituting (\ref{eq:BayesDetail1}), (\ref{eq:BayesDetail2}), and (\ref{eq:prior1}) into (\ref{eq:Bayes}), the posterior mean and variance of $x_{n,1}$, $\forall n\in[N]$, can be evaluated as
\begin{align}
\label{eq:postmean1}{\widehat x}_{n,1}&\!\!=\!\!\sum\nolimits_{x_{n,1}\in[K_a]}x_{n,1} p(x_{n,1}| r_{n,1})d x_{n,1},\\
\label{eq:postvar1}{\widehat v}_{n,1}&\!\!=\!\!\sum\nolimits_{x_{n,1}\in[K_a]}|x_{n,1}|^2 p(x_{n,1}| r_{n,1})d x_{n,1} - |{\widehat x}_{n,1}|^2.
\end{align}

\begin{prop}
\label{Proppp1}
    Substituting (\ref{eq:BayesDetail1}), (\ref{eq:BayesDetail2}), and (\ref{eq:prior2}) into (\ref{eq:Bayes}), the posterior distribution of $x_{n,m}$, $m\in\{2,3,...,M\}$ can be reformulated as
\begin{align}\label{eq:post2}
p(x_{n,m}| r_{n,m})&=(1-{\pi}_{n,m})\delta(x_{n,m})\\ \nonumber
&~~~~~~\!+\!{\pi}_{n,m}{\cal CN}\left(x_{n,m};{\mu}_{n,m},{\tau}_{n,m}\right),
\end{align}
where we have
\begin{align}
\label{eq:post1-2}{\mu}_{n,m}&=\left({\mu}_0 \varphi_{n,m}+{\tau}_0 {r}_{n,m}\right) \big / \left(\varphi_{n,m}+{ r}_{n,m}\right),\\
\label{eq:post2-2}\tau_{n,m}&=\left({ \tau}_0\varphi_{n,m}\right) \big /\left(\varphi_{n,m}+{r}_{n,m}\right),\\
\label{eq:post3-2}{{\cal L}}_{n,m}&={\rm ln}\dfrac{\varphi_{n,m}}{{\tau}_0+\varphi_{n,m}}-\dfrac{\left( {r}_{n,m}-{\mu}_0\right)^2}{{\tau}_0+\varphi_{n,m}}+\dfrac{\left| {{r}_{n,m}} \right|^2}{\varphi_{n,m}},\\
\label{eq:post4-2}{\pi}_{n,m}&=a_{n,m} \big /\left[{a_{n,m}+(1-a_{n,m}){\rm exp}\left(-{{{\cal L}}_{n,m}} \right)}\right].
\end{align}
\end{prop}
\renewcommand\qedsymbol{$\blacksquare$}
\begin{proof}\label{Proof:Theorem1}
Please refer to the Appendix \ref{proofP1}.
\end{proof}

Furthermore, according to (\ref{eq:post2}), we can obtain the posterior mean and variance of $x_{n,m}$, $m\in\{2,3,...,M\}$, respectively, denoted by
\begin{align}
\label{eq:postmean2}{\widehat x}_{n,m}&={ \pi}_{n,m}{\mu}_{n,m},\\
\label{eq:postvar2}{\widehat v}_{n,m}
&={\pi}_{n,m}\left( \left| {\mu}_{n,m}\right|^2+{\tau}_{n,m}\right)-\left( {\widehat x}_{n,m}\right)^2.
\end{align}

\vspace{-4mm}
\subsection{Parameter Estimation}
\subsubsection{Estimation of $K_a$}
According to (\ref{eq:postmean1}), we can obtain the estimated equivalent transmit signal vector $[\widehat{\bf X}_d]_{:,1}$, where the superscript $d$ denotes the $d$-th block, $\forall d\in[D]$. In addition, as indicated in (\ref{SpaX}), $\|[\widehat{\bf X}_d]_{:,1}\|_1$ should equal to $K_a$, in the case of perfect detection. To improve the robustness, we propose to acquire $\widehat{K}_a$ as follows
\begin{align}\label{eq:KaEst}
    \widehat{K}_a \!=\! \Psi\left\{\left\lfloor\left\|[\widehat{\bf X}_d]_{:,1}\right\|_1 \!\!+\! \frac{1}{2}\right\rfloor, d\in[D]\right\},
\end{align}
where function $\Psi\left\{\cdot\right\}$ calculates the most frequently occurring element of its argument. The intuition of (\ref{eq:KaEst}) is from majority voting \cite{Kaibin}, which improves the detection accuracy of $K_a$. We note that once $\widehat{K}_a$ is estimated, it is possible to go back and update the estimates of those blocks for which the estimated $\widehat{\bf X}_d$ did not correspond to $\widehat{K}_a$ active devices. We do not carry out this refinement in this paper.

\subsubsection{Estimation of the activity indicators}
The unknown parameter, denoted by $\bm{\theta}=\{a_{n,m}, \sigma^2, \tau_0, \mu_0, n\in[N],m\in[M]\}$, can be obtained by using the expectation maximization (EM) algorithm \cite{PRML}. The EM algorithm is an iterative approach
that effectively finds the maximum likelihood solutions for probabilistic models with unknown parameters \cite{PRML}. Here, the EM update rules are as follows
\begin{align}\label{eq:EM} 
\bm{\theta}^{i+1}={\rm arg}\max\limits_{\bm{\theta}}\mathbb{E}\left\{{{\rm ln}~p\left({{\bf X, Y}}\right)|{\bf Y};\bm{\theta}^i}\right\},
\end{align}
where $\bm{\theta}^i$ denotes the unknown parameter in the $i$-th iteration, $\mathbb{E}\{ \cdot|{\bf Y};\bm{\theta}^i\}$ represents the expectation conditioned on the received signal ${\bf Y}$ under $\bm{\theta}^i$. 

For example, to estimate the activity indicator $a_{n,m}$, $\forall n,m$, we partially differentiate $\mathbb{E}\left\{{{\rm ln}~p\left({{\bf X, Y}}\right)|{\bf Y}}\right\}$ with respect to $a_{n,m}$, and let it equal zero \cite{PRML}. According to (\ref{eq:Bayes}), the posterior distribution $p\left(x_{n,m}| [{\bf Y}]_{:,m}\right)$ can be approximated as $p\left(x_{n,m}| r_{n,m}\right)$, thereby simplifying the complexity of EM estimation. Hence, according to (\ref{eq:prior1}) and (\ref{eq:prior2}), the sparsity indicators $a_{n,1}$ and $a_{n,m}$, $m\in\{2,3,...,M\}$, can be obtained as follows:
\begin{align}
    \label{eq:act1}(a_{n,1})^{i+1}&=\sum\nolimits_{x_{n,1}\in[K_a]}p\left( x_{n,1}|r_{n,1};(a_{n})^i\right),\\
    \label{eq:act2}(a_{n,m})^{i+1}&=\pi_{n,m},
\end{align}
respectively. To exploit the row sparsity as indicated in (\ref{SparseMIMO}), we further average the activity indicators on $M$ antennas as
\begin{align}
    \label{eq:actT}(a_n)^{i+1}=\frac{1}{M}\sum\nolimits_{m=1}^M (a_{n,m})^{i+1}, \forall n.
\end{align}

\subsubsection{Estimation of other parameters}
Furthermore, we apply the EM update rules (\ref{eq:EM}) to estimate other unknown parameters. In particular, to estimate the noise variance, we partially differentiate $\mathbb{E}\left\{{{\rm ln}~p\left({{\bf X, Y}}\right)|{\bf Y}}\right\}$ with respect to $\sigma^2$ and let it equal zero \cite{PRML}. Similarly, calculations can be done to obtain $\tau_0$ and $\mu_0$. Note that only (\ref{eq:prior2}) includes $\tau_0$ and $\mu_0$, suggesting that the messages from the first antenna do not contribute to the estimation of $\tau_0$ and $\mu_0$. Finally, we obtain the EM update rules of the noise variance $\sigma^2$, the unknown $\tau_0$, and $\mu_0$ as $({\sigma^2})^{i+1}\!\!=\!\!\frac{1}{MN}\!\!\!\sum\limits_{l,m=1}^{L,M}\!\![{\frac{({[{{\bf Y}}]_{l,m}\!\!-\!(Z_{l,m})^i})^2}{({1+\frac{(V_{l,m})^i}{({\sigma^2})^i}})^2}+\frac{({\sigma^2})^i (V_{l,m})^i}{(V_{l,m})^i+({\sigma^2})^i}}]$, $(\tau_0)^{i+1}\!\!=\!\!\frac{\sum\nolimits_{m=2}^{M}\sum\nolimits_{n=1}^{N}(\pi_{n,m})^{i}\![((\mu_0)^{i+1}-({\mu}_{n,m})^{i} )^2+({\tau}_{n,m})^{i} ]}{\sum\nolimits_{m=2}^{M}\sum\nolimits_{n=1}^{N}(\pi_{n,m})^{i}}$, and $(\mu_0)^{i+1}=\frac{\sum\nolimits_{m=2}^{M}\sum\nolimits_{n=1}^{N}(\pi_{n,m})^{i}({\mu}_{n,m})^{i}}{\sum\nolimits_{m=2}^{M}\sum\nolimits_{n=1}^{N}(\pi_{n,m})^{i}}$, respectively.

\begin{algorithm}[t]
\algsetup{linenosize=\small} \small
\caption{Proposed AMP-DA Algorithm}\label{Algorithm:2}
\begin{algorithmic}[1] 
\raggedright 
\REQUIRE Received signals ${\bf Y}_{d}\in \mathbb{C}^{L\times M}$, modulation codebook ${\bf P}\in\mathbb{C}^{L\times N}$, quantization codebook ${\bf U}\in\mathbb{R}^{Q\times N}$, maximum iteration number $I_0$, and damping parameter $\tau$.
\ENSURE Estimated global model update $\widehat{\bf s}$.
\STATE ${\forall d, n, l}$: Initialize iteration index $i$~$=$~1, sparsity indicators $(a^{d}_n)^1=0.5$, noise variance $(\sigma^2)^1=100$, residual $R^0=100$, $(Z^{d}_{l,m})^0=\left[{\bf Y}_{d}\right]_{l,m}$, $(V^{d}_{l,m})^0=1$, $({\widehat x}_{n,m}^{d})^1=0$, and $({\widehat v}_{n,m}^{d})^1=1$;
\label{A1:initial}
\FOR {$i=1$ to $ I_0$}
\label{A1:T0}
\STATE  Compute $(V^{d}_{l,m})^i$, $(Z^{d}_{l,m})^i$, $(\varphi^{d}_{n,m})^i$, $(r^{d}_{n,m})^i$  using (\ref{eq:UpdateSigma})-(\ref{eq:UpdateZ}); 
\\\%\textbf{Decoupling steps}
\label{A1:decoupling}
\STATE Damping: $(V^{d}_{l,m})^i=\tau(V^{d}_{l,m})^{i-1}+(1-\tau)(V^{d}_{l,m})^i$, \\
$(Z^{d}_{l,m})^i=\tau(Z^{d}_{l,m})^{i-1}+(1-\tau)(Z^{d}_{l,m})^i$;
\label{A1:damping}
\\ {\% \textbf{Denoising steps}}
\STATE Compute $({\widehat x}_{n,1}^{d})^i$ and $({\widehat v}_{n,1}^{d})^i$ using (\ref{eq:postmean1}) and (\ref{eq:postvar1});
\label{A1:denoising1}
\STATE Compute $({\widehat x}_{n,m}^{d})^i$ and $({\widehat v}_{n,m}^{d})^i$ using (\ref{eq:postmean2}) and (\ref{eq:postvar2}), $m\in\{2,...,M\}$;
\label{A1:denoising2}
\\ \textbf{\textbf{\%}Parameter update:}
\label{A1:EM-S}
\STATE Update $(a^{d}_n)^i$ using (\ref{eq:actT}), update $(\sigma^{2})^i$, $(\mu_0^d)^{i}$, and $(\tau_0^d)^{i}$;
\label{A1:EM-ACT}
\STATE Calculate residual: $R^i = \dfrac{1}{LD}\sum\limits_{d\in[D]}\left\| {\bf Y}_{d}-{\bf P}(\widehat{\bf X}_{d})^{i} \right\|_F$;
\label{A1:Res}
\STATE $i = i+ 1;$
\IF{$i>15$ \text{and} $R^{i+1}\geq R^i$}
\label{A1:if}
\STATE ${\bf break}$;~~\{End iterations\}
\label{A1:break}
\ENDIF
\label{A1:endif}
\ENDFOR
\label{A1:endfor}
\STATE ${\forall d}$:Estimated equivalent transmit signal $\widehat{\bf x}_{d}=[(\widehat{\bf X}_{d})^i]_{:,1}$;
\label{A1:Xest}
\\ \textbf{\textbf{\%}Estimate the number of active devices:}
\STATE Estimate $\widehat{K}_a$ using (\ref{eq:KaEst});
\label{A1:KaEst}
\\ \textbf{\textbf{\%}Model aggregation:}
\STATE Obtain global model update $\widehat{\bf s}=\frac{1}{\widehat{K}_a} \left({\bf I}_D\otimes{\bf U}\right) \left[ \widehat{\bf x}_1^T,...,\widehat{\bf x}_D^T  \right]^T$.
\label{A1:DA}
\end{algorithmic}
\end{algorithm}

\subsection{Proposed AMP-DA Algorithm}
Based on (\ref{eq:UpdateSigma})-(\ref{eq:actT}), we summarize the proposed AMP-DA algorithm in {\bf Algorithm} \ref{Algorithm:2}. The details are explained as follows. 
In line \ref{A1:initial}, we initialize the sparsity indicators $a^{d}_n$, the variables $V^{d}_{l,m}$, $Z^{d}_{l,m}$, the posterior mean ${\widehat x}_{n,m}^{d}$, and posterior variance ${\widehat v}_{n,m}^{d}$, $\forall d, n, l,m$. The iteration starts in line \ref{A1:T0}. Specifically, lines \ref{A1:decoupling}-\ref{A1:denoising2} correspond to the AMP operation. In the $i$-th iteration of the AMP decoupling step (line \ref{A1:decoupling}), $V^{d}_{l,m}$, $Z^{d}_{l,m}$, $\varphi_{n,m}^{d}$, and $r_{n,m}^{d}$, are calculated according to (\ref{eq:UpdateV}), (\ref{eq:UpdateZ}), (\ref{eq:UpdateSigma}), and (\ref{eq:UpdateR}), respectively. Furthermore, a damping parameter $\tau$ is adopted in line \ref{A1:damping} to prevent the algorithm from diverging \cite{M-AMP}. In addition, in the AMP denoising steps (line \ref{A1:denoising1} and line \ref{A1:denoising2}), we calculate the posterior mean ${\widehat x}_{n,1}^{d}$ and the corresponding posterior variance ${\widehat v}_{n,1}^{d}$, of the $i$-th iteration, by using (\ref{eq:postmean1}) and (\ref{eq:postvar1}), respectively. In addition, for $m\in\{2,...,M\}$, ${\widehat x}_{n,m}^{d}$ and ${\widehat v}_{n,m}^{d}$ are calculated by using (\ref{eq:postmean2}) and (\ref{eq:postvar2}). Accordingly, EM operation is used to update the sparsity indicators $a^{d}_n$, the noise variance $\sigma^{2}$, and the mean and variance of the prior distribution $\mu_0^d$ and $\tau_0^d$, respectively, in line \ref{A1:EM-ACT}. The residual of the $i$-th iteration is calculated in line \ref{A1:Res}. Then, the iteration restarts in line \ref{A1:decoupling} until the maximum iteration number $I_0$ is reached. Otherwise, if line \ref{A1:if} is triggered, i.e., the residual does not goes down after at least 15 steps, we stop iterations. After the iteration, we acquire the estimated equivalent transmit signal $\widehat{\bf x}_{d}=[(\widehat{\bf X}_{d})^i]_{:,1}$, ${\forall d}$, in line \ref{A1:Xest}. According to (\ref{eq:KaEst}), in line \ref{A1:KaEst}, we estimate the number of active devices $\widehat{K}_a$ based on $\widehat{\bf x}_{d}$, ${\forall d}$. Finally, according to (\ref{computingTask}) and Section III-B, we obtain the global model update $\widehat{\bf s}$ in line \ref{A1:DA} by using $\widehat{\bf x}_{d}$ and $\widehat{K}_a$, ${\forall d}$, and the weight of the ML model will be updated using $\widehat{\bf s}$.

\begin{remark}
Since only $[\widehat{\bf X}_{d}]_{:,1}$ matters for the computing task (\ref{computingTask}), the digital model aggregation can be done by only using the received signal at the first antenna of the BS, i.e., running Algorithm \ref{Algorithm:2} except line \ref{A1:denoising2}. If multiple antennas are available at the BS, the proposed AMP-DA algorithm improves the estimation accuracy of $[{\bf X}_{d}]_{:,1}$ by exploiting the row sparsity, as described in (\ref{SparseMIMO}). As a result, the training accuracy of the proposed MD-AirComp improves with multi-antenna BS, which will be verified in the simulations.
\end{remark}

\subsection{Computational Complexity}
The computational complexity of the proposed AMP-DA algorithm mainly comes from the complex-valued matrix multiplications, i.e., (\ref{eq:UpdateSigma})-(\ref{eq:UpdateZ}), (\ref{eq:postmean1}), (\ref{eq:postvar1}), (\ref{eq:postmean2}), and (\ref{eq:postvar2}), at each iteration. The overall computational complexity at one FEEL training round is of the order of ${\cal O}(I_0LNMD)$, which equals to ${\cal O}(I_0LM2^JW/Q)$. It can be seen that the overall complexity scales linearly with the quantization level $N$, which is appealing for efficient detection with high-dimensional VQ along with a large quantization level. {The computational complexity of our proposed AMP-DA algorithm can be reduced further by employing deep unfolding schemes \cite{Unfolding}, which is left as a future research direction.}

\section{Convergence Analysis}
In this section, we present the theoretical convergence results of our proposed MD-AirComp. We first introduce some assumptions needed for the proof. For simplicity, $[\widehat{\bf X}_{d}]_{:,1}$ is denoted by ${\bf x}_{d}$ in this and the following sections.

\begin{assumption}[Smoothness]\label{as:smooth}
Each loss function on the $k$-th device, $F_k({\cdot})$ is $\overline{L}$-smooth, i.e., $\forall \wb,\overline{\wb} \in \RR^W$, 
\vspace{-5pt}
\begin{align}
    \left|F_k(\wb)-F_k(\overline{\wb})-\langle\nabla F_k(\overline{\wb}), \wb-\overline{\wb}\rangle\right| \leq \frac{\overline{L}}{2}\|\wb-\overline{\wb}\|^2.
\end{align}
This indicates the $\overline{L}$-gradient Lipschitz condition, i.e., $\|\nabla F_k(\wb) - \nabla F_k(\overline{\wb}) \| \leq \overline{L} \|\wb - \overline{\wb} \|$, where $\overline{L}$ is a constant.
\end{assumption}

\begin{assumption}[Bounded Gradient]\label{as:bounded-g}
Each loss function on the $k$-th device $F_k(\wb)$ has $G$-bounded stochastic gradient on $\ell_2$ norm, i.e., for all ${\bm \varepsilon}$, we have $\|\nabla F_k(\wb,\xi)\|\leq G$.
\end{assumption}

\begin{assumption}[Bounded Variance]\label{as:bounded-v} Each stochastic gradient on the $k$-th device has a bounded local variance, i.e., for all $\wb, k \in [K]$,we have
    $\EE \left[ \|\nabla F_k(\wb,{\bm \varepsilon})- \nabla F_k(\wb)\|^2\right] \leq \sigma_l^2$,
and the loss function on each device has a global variance bound,
$\frac{1}{K}\sum_{k=1}^K \|\nabla F_k(\wb)-\nabla f(\wb)\|^2 \leq \sigma_g^2$.
\end{assumption}
Note that ${\bm \varepsilon}$ denotes a random local data sample. The bounded local variance represents the randomness of stochastic gradients, and the bounded global variance represents data heterogeneity among devices. Specifically, $\sigma_g = 0$ corresponds to the i.i.d setting, in which the datasets of all the clients have the same distribution. Assumptions \ref{as:smooth}-\ref{as:bounded-v} are widely adopted in federated optimization problems \cite{FLqChang, FedPAQ,FedAdam,FedAMS}.

\begin{assumption}[Biased Quantizer]\label{as:compressor}Consider a biased operator $\cC: \RR^Q \to \RR^Q$: $\forall \zb \in \RR^Q$, there exists a constant $0\leq q \leq 1$ such that 
\begin{align}
    \|\cC(\zb)-\zb\| \leq q \|\zb\|, \forall \zb\in \RR^Q.
\end{align}
\end{assumption}

Assumption \ref{as:compressor} is a standard assumption for biased compressors, which was adopted to model scaled-sign and top-$k$ compressors in the literature. For such biased quantizer, there exists a constant $\gamma$ such that, for global training round $t \in [T]$, we have 
\begin{align}\label{Quant_Dissi}
\resizebox{0.97\linewidth}{!}{$
\left\|\cC\left( \frac{1}{K}\!\! \sum_{k=1}^K \left[\Delta_k^t + \eb_k^t\right] \right) \!\!- \!\frac{1}{K} \!\!\sum_{k=1}^K \cC\left(\Delta_k^t + \eb_k^t\right)\right\|   \!\!\leq\! \gamma \left\| \frac{1}{K}\!\! \sum_{k=1}^K \Delta_k^t \right\|,
$}
\end{align}
which bounds the difference between the average of the quantized values and the quantized value of the averages \cite{FedAMS}.

\begin{assumption}[AMP Estimation]\label{as:AMPasumption}We assume that the number of non-zero elements in $\xb_d$, i.e., $\|{\bf x}_d\|_0$, $\forall d\in[D]$, is smaller than the length of the modulation codeword $L$. This enables AMP algorithm to reconstruct $\xb_d$ with an estimation MSE of $\EE\|\widehat{\xb}_d - \xb_d\|^2=N\sigma_n^2$, $\forall d$ \cite{Donoho}. In addition, we can assume the perfect estimation of the number of active devices $K_a$.
\end{assumption}

As described in (\ref{SpaX}), $\|{\bf x}_d\|_0\leq K_a$. In practice, the communication system can have prior knowledge about the maximum number of active devices $K_a$ at each iteration, and can select an $L$ that satisfies Assumption \ref{as:AMPasumption}. 
Note that this assumption simplifies the analysis, while we will describe how the estimation error affects the performance later.

In the following, we will show the convergence results\footnote{For simplicity, we only present the convergence guarantee for full participation of devices. Note that the theoretical analysis can be easily extended to partial participation according to \cite{FedAMS}.} of the proposed MD-AirComp-based FEEL scheme.

\begin{theorem}\label{thm:full-noncomp}
Under Assumptions~\ref{as:smooth}-\ref{as:AMPasumption}, if the local learning rate $\eta_l$ satisfies the following condition: $\eta_l \leq \min \left\{\frac{1}{8K\overline{L}}, \frac{1}{\eta \overline{L} {T_l} (3+2C_1)}\right\}$, then the iterates of MD-AirComp-based FEEL satisfy
\begin{align}\label{eq:conve-thm}
    \min_{t\in [T]} \EE [\|\nabla f(\wb^t)\|^2] \leq 4 \left[\frac{f_0-f_*}{\eta \eta_l {T_l} T} + \Xi + \Psi \right],
\end{align}
where $f_0\triangleq f(\wb^0)$, $f_*\triangleq {\rm min}_{\wb}f(\wb)$, $\Xi \triangleq (\frac{3}{2}+C_1) \eta \overline{L} \sigma_l^2 \frac{ \eta_l}{K} + \frac{5\eta_l^2 {T_l} \overline{L}^2 }{2} (\sigma_l^2+6{T_l} \sigma_g^2), \Psi \triangleq \frac{\eta\eta_l \overline{L}C_0T_lG^2N^2 \sigma_n^2}{K^2}$, $C_1 \triangleq \frac{2(q+\gamma)^2}{(1-q^2)^2}$, and $C_0\triangleq\frac{2(1+q^2)^2}{(1-q^2)^2}$.
\end{theorem}

\renewcommand\qedsymbol{$\blacksquare$}
\begin{proof}\label{Proof:Theorem1}
Please refer to the Appendix \ref{proofT1}.
\end{proof}

\begin{remark}\label{ConvAna}
It can be seen that the upper bound for $\min_{t\in [T]} \EE [\|\nabla f(\wb_t)\|^2]$ contains three terms: 1) The first term, is due to the initialization error, i.e. $f_0-f_*$. 2) The second term $\Xi$ relates to the local stochastic variance $\sigma_l$ and global variance $\sigma_g$. In the i.i.d setting, we have zero global variance, i.e., $\sigma_g = 0$, and $\Xi$ will be smaller and less dependent on the number of local training steps $T_l$. Note that $C_1$ and $C_0$ are two constants related to the quantization error. If the quantization error increases, i.e., $q$ ($0<q<1$) increases, $C_1$ and $C_0$ will increase as well. 3) Finally, the third term $\Psi$ is due to the wireless communication errors, i.e., $\| \widehat{\bf x}_d^{t}\!-\!{\bf x}_d^{t}\|^2$, $\forall t,d$. Note that if the wireless channel is noiseless, we have $\Psi=0$. In practice, a lower detection MSE at the BS will reduce $\Psi$, resulting in a tighter convergence bound. 
\end{remark}

\begin{corollary}\label{cor:fedams-full}
Suppose we choose the global learning rate $\eta= \Theta(\sqrt{K})$ and local learning rate $\eta_l= \Theta(\frac{1}{T_l\sqrt{T}})$, when $T$ is sufficiently large. With this choice of the learning rates, the convergence rate for MD-AirComp-based FEEL is independent of the communication errors and quantization effects, and satisfies
\begin{align}\label{eq:fedams-full-cor}
    & \min_{t\in [T]} \EE \left[\|\nabla f(\wb_t)\|^2\right] =  \cO\left(\frac{1}{\sqrt{T}}\right).
\end{align}
\end{corollary}

\section{Simulation Results}

\subsection{Parameter setting}
In this section, we evaluate the performance of the proposed MD-AirComp scheme by considering FEEL for the image classification task on the commonly adopted CIFAR-10 dataset \cite{FL-IoT}. The CIFAR-10 dataset has a training set of 50,000 and a test set of 10,000 colour images of size 32 × 32 in 10 classes. We consider $K=40$ devices (unless otherwise specified) in the cell collaboratively training a neural network (the structure follows the ResNet \cite{ResNet}) with 269722 (i.e., $W \approx 2.7\times10^5$) parameters. We consider non-i.i.d. local training data across the devices as follows \cite{Yulin}: 1) Each one of the 40 devices is randomly assigned 250 samples from the dataset; 2) The remaining 40,000 samples are sorted by their labels and grouped into 40 shards of size 1000. Then, each device is assigned one shard\footnote{Note that the earth mover’s distance (EMD) for i.i.d. local datasets equals to zero \cite{FL_NonIID}. In our simulation, we have $\text{EMD}=1.44$, which represents highly non-i.i.d. local datasets \cite{FL_NonIID}.}. Furthermore, we set the number of local training iterations to $E=3$, the mini-batch size to 20, the local learning rate to $\eta_l=0.01$, and the global learning rate to $\eta=1$. The metric to evaluate the performance is the {\it test accuracy} of the model on the test dataset.

\begin{figure}[t] 
\vspace{-2mm} 
\centering  
\includegraphics[width = 0.66\columnwidth]{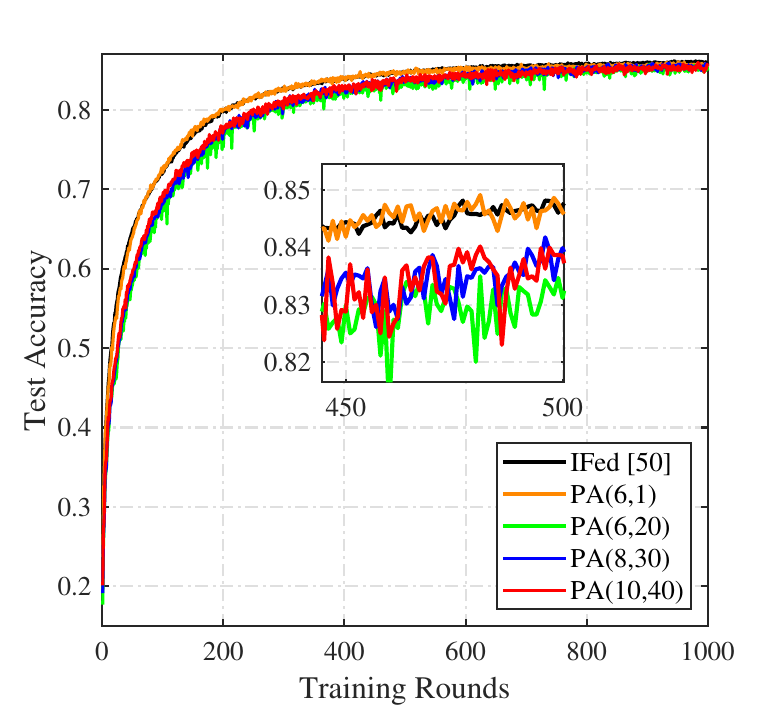}  
\captionsetup{font={footnotesize, color = {black}}, singlelinecheck = off, justification = raggedright,name={Fig.},labelsep=period}
\caption{Test accuracy vs training rounds under different VQ parameters.}  
\vspace{-3mm}
\label{TA_quant}
\end{figure}

We adopt the K-means++ algorithm for VQ to learn the quantization codebook \cite{KmeansPP}. Since the distribution of the model updates changes from one training round to the next \cite{FedDQ}, we also update the VQ codebook to achieve better performance. We assume that the BS has available a small local dataset and performs local training as well. In each communication round, the BS obtains a local model update with error accumulation ${\bf s}'\in\mathbb{R}^W$ as described in Section III-A. Then the BS uses its own local update ${\bf s}'$ to determine the VQ codebook. In particular, the $D$ data samples $\{\left[{\bf s}'\right]_{(d-1)Q+1:dQ}\}_{d=1}^D$ are used by the VQ algorithm to obtain the quantization codebook ${\bf U}\in\mathbb{R}^{Q\times N}$, where $N=2^J$ is the number of clusters. Then, the BS broadcasts\footnote{Note that the number of real values in ${\bf U}$ is $NQ$. Considering $N=256$ and $Q=20$, we can obtain $\frac{NQ}{W}\approx1.9\%$, which implies that the communication cost of broadcasting $\bf U$ is very small in comparison to broadcasting the updated model parameters.} codebook ${\bf U}$ to the devices together with the global model. Note that we pad zeros at the end of ${\bf s}'$ to ensure that ${\bf s}'$ is divisible by $Q$. At the receiver, those padded elements will be deleted. Since $W$ is much larger than $Q$, the effects of padding can be negligible. Different quantization parameters $J$ and $Q$ are tested in Fig. \ref{TA_quant}. We remark that the VQ quantizer is a biased quantizer as prescribed by Assumption \ref{as:compressor}.

As for the wireless transmission, we assume that 30\% of the devices are active in each round. {Since the BS does not exactly know $K_a$, we use $0.4\times K$ instead of the actual $K_a$ in (\ref{eq:prior1}) as a prior information.} We set $\text{SNR}=20$\,dB unless otherwise specified. For Algorithm 2, the damping parameter is set to $\tau=0.3$, and the maximum iteration number to $I_0=50$. The elements of modulation codebook ${\bf P}\in\mathbb{C}^{L\times N}$ obey i.i.d. complex Bernoulli distribution, which satisfies the restricted isometry property and is commonly adopted in CS \cite{Qiao2}. {In addition, as mentioned in Remark \ref{CSIconditionBad}, to avoid extremely small SNR at the first receive antenna, we set $\epsilon_h=0.14$ as the threshold of pre-equalization. We note that in practice, this threshold can be determined by the transmit power constraint.}

For comparison, we consider the following schemes. {\bf IFed}: IFed is the acronym for \textit{ideal federated learning}, where the communication channels are assumed ideal and there is no quantization distortion. The IFed algorithm adopted for the simulations is FedAvg\footnote{Note that all the schemes are built upon FedAvg in our simulation for the sake of fairness, while FedAvg can be replaced by other optimization algorithms, e.g., adaptive federated optimizer (FedAdam) \cite{FedAdam}, for better convergence.} \cite{FedAvg}. {\bf PA ($J$, $Q$)}: Here, PA is the acronym for \textit{perfect aggregation} at the receiver, while the transmitter is the same as the proposed MD-AirComp scheme. The number of quantization bits and the VQ dimension are denoted by ($J$, $Q$). {\bf MD-AirComp ($J$, $Q$, $L$)}: The proposed MD-AirComp scheme, where the number of quantization bits, the VQ dimension, and the length of modulation codewords are denoted by ($J$, $Q$, $L$). Different numbers of antennas $M$ at the BS are also tested in the simulations. {\bf OBDA}: The state-of-the-art digital AirComp scheme with one-bit quantization introduced in \cite{Kaibin}. Note that the main parameters of the OBDA scheme are adopted as released in \cite{Kaibin}. {However, we observed that learning rate has a significant impact on the convergence of OBDA, especially when dealing with non-i.i.d. local training data. To achieve a better test accuracy of OBDA, we carefully choose the cosine annealing scheduler \cite{CosAnne} for learning rate adaptation, where the initial learning rate is 0.001, the minimal learning rate is 0.0001, and the cosine function follows a period of 1400.} {\bf FSK-MV}: FSK-MV is the acronym for {\it frequency shift keying-based majority vote} proposed in \cite{FSKMV}, which is the best performing FSK-based method in terms of communication overhead \cite{FSKMV,DigFL}. Based on SignSGD, the learning parameters of FSK-MV are the same as OBDA. At the transmitters, each binary element of the model update vector is transmitted by sending a fixed power pilot signal on either of the two pre-allocated subcarrier frequencies. No pre-equalization needed at the transmitter. The receiver detects the signal power on the two subcarriers for majority voting.

\subsection{Communication Overhead}
Furthermore, we compare the communication overhead of different schemes. We consider an orthogonal frequency-division multiplexing (OFDM) system, where $P$ subcarriers are available. Due to the large number of devices, the communication overhead is dominated by the UL communications. We consider the UL communication overhead of one training round, i.e., $K_a$ active devices quantize and transmit their model update ($W$ elements) to the BS. With one-bit quantization, the number of symbols transmitted by OBDA\footnote{Considering binary phase shift keying (BPSK) symbols as used in \cite{Kaibin}.} and FSK-MV is $W$ and $2W$, respectively. It can be seen that FSK-MV does not require CSI at a scarifies of larger communication overhead. Furthermore, the proposed MD-AirComp transmits $\lceil\frac{W}{Q}\rceil L$ symbols \footnote{In our conference paper \cite{ISIT}, real-value modulation codebook ${\bf P}\in\mathbb{R}^{L\times N}$ was considered. In this case, if using I/Q channels, transmitting $L$-real values only costs $\lceil L/2 \rceil$ symbol. While, in this work, we extend the setting to a more general case, i.e., ${\bf P}\in\mathbb{C}^{L\times N}$.}. Our simulation results in the following show that $L$ can be set smaller than $Q$, making the communication overhead of MD-AirComp smaller than that of the OBDA and FSK-MV schemes. In addition, for the above digital AirComp schemes, since all the devices share $P$ subcarriers, the number of time slots needed equals to the number of transmitted symbols divided by $P$.

As for conventional digital communication, we consider the orthogonal frequency-division multiple access (OFDMA) and the quadrature amplitude modulation (QAM). For fair comparison, the same VQ at the transmitter is considered. After quantization, the index of each block ($J$ bits) is modulated to a QAM symbol. Hence, each device needs to transmit $\lceil\frac{W}{Q}\rceil$ symbols. We assume each device is uniformly allocated with $\frac{P}{K}$ subcarriers. Hence, the number of time slots needed is $\lceil\frac{W}{Q}\rceil\frac{K}{P}$. We denote the conventional digital communication scheme as `VQ + OFDMA', and we compare the number of time slots (i.e., transmission delay), of different schemes in Table \ref{CommOverhead}. As mentioned in Remark \ref{UEspar}, $K_a \ll K$, while $L$ has values close to, or possibly smaller (due to devices collision) than, $K_a$, the transmission delay of MD-AirComp is obviously far smaller than the `VQ + OFDMA' scheme. Numerical results in Table \ref{CommOverhead} verify the lower communication overhead of the proposed MD-AirComp scheme. Note that the best performance that ``VQ + OFDMA" scheme can achieve is the PA benchmark, if considering perfect QAM demodulation.

\begin{table}[!t]\small
\centering
\captionsetup{font = {normalsize}, labelsep = period} 
\begin{threeparttable}[b]
\caption{UL communication overhead comparison}
\begin{tabular}{|c|c|c|}
\Xhline{1.2pt}
Schemes & Number of Time Slots & Numerical Values\tnote{1} \\
\Xhline{1.2pt}
VQ + OFDMA & $\big\lceil\frac{W}{Q}\big\rceil\frac{K}{P}$ & 527 \\
\hline
FSK-MV \cite{FSKMV} & $\frac{2W}{P}$ & 527 \\
\hline
OBDA \cite{Kaibin} & $\frac{W}{P}$ & 264 \\
\hline
MD-AirComp & $\big\lceil\frac{W}{Q}\big\rceil\frac{L}{P}$ & \makecell{264 ($L=20$)\\198 ($L=15$)} \\
\Xhline{1.2pt}
\end{tabular}
\begin{tablenotes}
\item[1] These values are calculated and rounded up, referring to the simulation parameters, i.e., $W=269722$, $Q=20$, $L=20$ (or $L=15$) and $K=40$. The number of subcarriers $P=1024$.
\end{tablenotes}
\label{CommOverhead}
\end{threeparttable}
\end{table}

\subsection{Results of FEEL}
In Fig. \ref{TA_quant}, we present how the test accuracy changes with the training rounds for IFed and PA ($J, Q$) for different $J$ and $Q$ values. We observe that scalar quantization (i.e., $Q=1$) with $J=6$ achieves nearly the same test accuracy as the IFed benchmark. In addition, the test accuracy under other VQ parameters nearly overlap. Note that when the VQ dimension $Q$ increases, we need more quantization bits $J$ to convey the corresponding update with reasonable accuracy \cite{PRML}. In the remaining simulations, we set $J=6$ and $Q=20$ to test the performance of the proposed MD-AirComp scheme.

\begin{figure}[t]
 \vspace{-6mm}
\centering
\subfigure[]{
    \begin{minipage}[t]{1\linewidth}
        \centering
         \includegraphics[width=0.66\columnwidth]{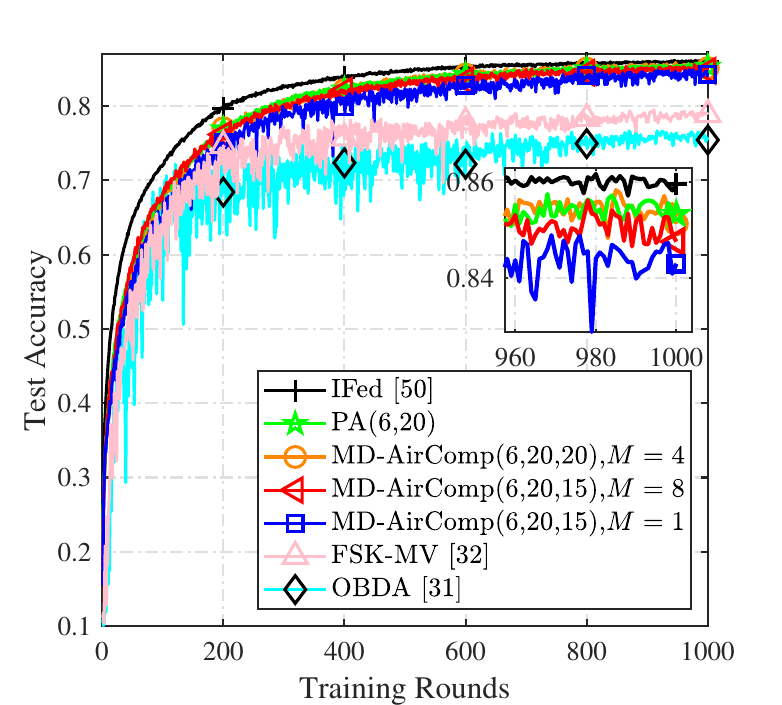}\\
\label{FigTAa}
        
    \end{minipage}%
}\\
\vspace{-3.5mm}
\subfigure[]{
    \begin{minipage}[t]{1\linewidth}
        \centering
        \includegraphics[width=0.66\columnwidth]{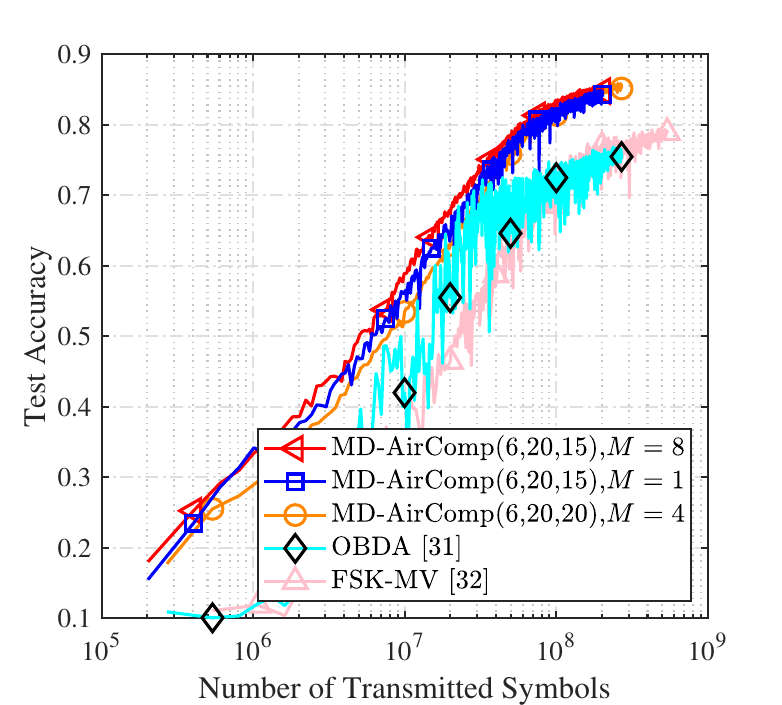}\\
\label{FigTAb}
    \end{minipage}%
}%
\centering
\setlength{\abovecaptionskip}{-1mm}
\captionsetup{font={footnotesize}, singlelinecheck = off, justification = justified,name={Fig. },labelsep=period}
\caption{FEEL performance comparison of the proposed MD-AirComp scheme with the benchmarks: (a) Test accuracy vs training rounds; (b) Test accuracy vs the number of transmitted symbols.}
\label{FigTA}
 \vspace{-5mm}
\end{figure}

As shown in Fig. \ref{FigTAa}, the test accuracy of the proposed MD-AirComp scheme converges to the IFed benchmark. We see slower convergence and some performance loss in the final test accuracy of OBDA and FSK-MV compared with MD-AirComp, which is expected considering the one-bit quantization loss in these schemes. FSK-MV  converges faster and improves the final accuracy compared with OBDA, while requiring a larger communication overhead. Furthermore, it can be seen that ``MD-AirComp (6, 20, 15), $M=1$" performs worse than the PA (6, 20) due to the communication error under limited observation $L$. Fortunately, by exploiting the multiple antennas at the BS, we can reduce the communication error without increasing the communication resources. For example, it is shown that ``MD-AirComp (6, 20, 15), $M=8$" achieves nearly the same accuracy as that of the PA (6, 20). From Fig. \ref{FigTAb}, it can be seen that the MD-AirComp scheme is more communication efficient than ever the OBDA scheme, that is, to achieve the same test accuracy it requires the transmission of fewer symbols over the channel. In addition, although the MD-AirComp (6, 20, 20) achieves slightly better test accuracy as shown in Fig. \ref{FigTAa}, it is less communication efficient than ``MD-AirComp (6, 20, 15), $M=8$". This reveals that we can reduce the code-length $L$, and hence, increase the bandwidth efficiency of MD-AirComp if more antennas are available at the receiver end.

\begin{figure}[t]
 \vspace{-6mm}
\centering
\subfigure[]{
    \begin{minipage}[t]{1\linewidth}
        \centering
        \includegraphics[width=0.66\columnwidth]{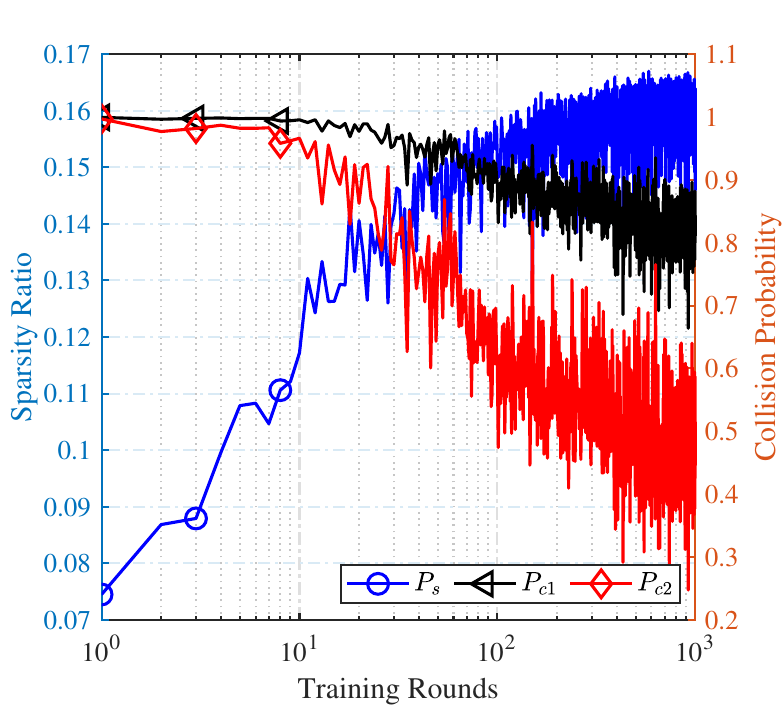}\\
\label{FigWTa}
    \end{minipage}%
}\\
\vspace{-3.5mm}
\subfigure[]{
    \begin{minipage}[t]{1\linewidth}
        \centering
        \includegraphics[width=0.66\columnwidth]{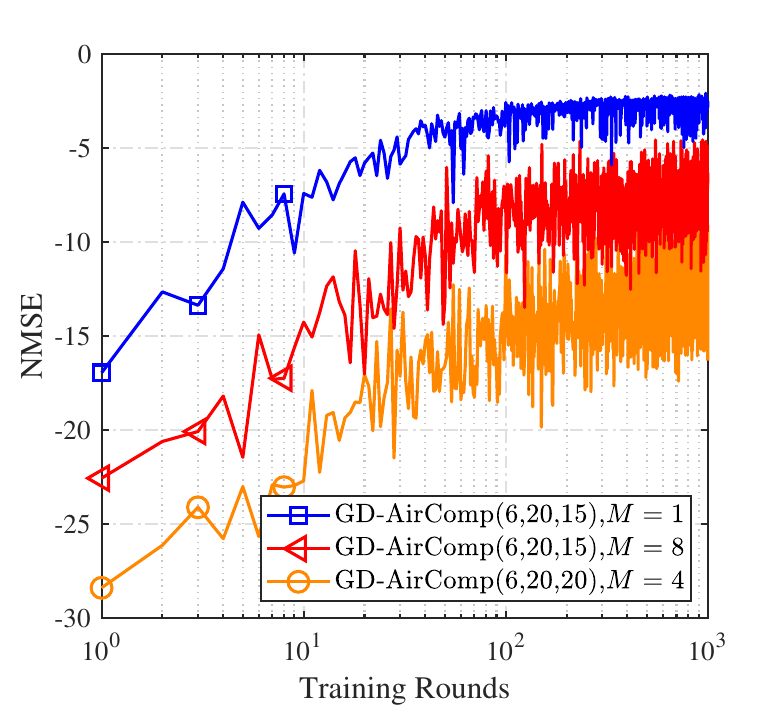}\\
\label{FigWTb}
    \end{minipage}%
}
\centering
\setlength{\abovecaptionskip}{-1mm}
\captionsetup{font={footnotesize}, singlelinecheck = off, justification = justified,name={Fig.},labelsep=period}
\caption{Wireless transmission performance of the proposed MD-AirComp scheme: (a) Sparsity ratio and device collision probability vs training rounds; (b) Detection NMSE vs training rounds.}
\label{FigWT}
 \vspace{-5mm}
\end{figure}

\vspace{-3mm}
\subsection{Results of Wireless Transmission}
In Fig. \ref{FigWTa}, we investigate the average sparsity ratio $P_s\triangleq\frac{\sum_{d=1}^D\|{\bf x}_d\|_0}{ND}$, and collision probability of MD-AirComp (6, 20, 20) versus the global training rounds\footnote{We observed that MD-AirComp (6, 20, 20) and MD-AirComp (6, 20, 15) have similar average sparsity ratio and collision probability during the training. For simplicity, we only present the related results of MD-AirComp (6, 20, 20).}. {We denote the probability of device collision occurred in $D$ blocks by $P_{c1}\triangleq 1-P\{d\in[D]: \|{\bf x}_d\|_0=K_a \}$. In addition, the probability that more than two devices collide is denoted by $P_{c2}\triangleq P_{c1}-P\{d\in[D]:\|{\bf x}_d\|_0=K_a-1 \}$. To enhance the clarity of the presentation, we employ logarithmic scaling on the x-axis.} It can be seen that with fewer device collisions, the sparsity ratio increases along training rounds. The reason is that the variance of model updates gradually decrease as the training progresses\cite{FedDQ}. Hence, with the same number of quantization levels, the values of different devices' model updates become more distinguishable, resulting in a decrease in the collision probability. 

For given number of observations $L$ and signal dimension $N$, the reconstruction performance of CS algorithms gets worse if the sparsity ratio of the signal increases, i.e., as the signal becomes less sparse. {The metric to evaluate the signal reconstruction performance is normalized MSE (NMSE), which is expressed as
\begin{align}
    \text{NMSE} = 10\log_{10}\left(\dfrac{\sum\nolimits_{d=1}^{D}\| [{\bf X}_{d}]_{:,1} \!\! - \!\! [\widehat{\bf X}_{d}]_{:,1} \|_2^2}{\sum\nolimits_{d=1}^{D}\| {\bf X}_{d}]_{:,1} \|_2^2}  \right).
\end{align}}

From Fig. \ref{FigWTb}, we can observe that NMSE decreases due to the increase in the sparsity ratio. In addition, NMSE of ``MD-AirComp (6, 20, 15), $M=8$" is better than that of ``MD-AirComp (6, 20, 1), $M=1$", which verifies the performance improvement shown in Fig. \ref{FigTAa}. Furthermore, NMSE of ``MD-AirComp (6, 20, 20), $M=4$" is the best due to more observations. Although NMSE of ``MD-AirComp (6, 20, 15), $M=8$" is smaller than -10\,dB when the training rounds is more than 200, its test accuracy is as good as the PA scheme due to the error correction property of FL. 

\begin{figure}[t] 
\vspace{-4mm} 
\centering  
\includegraphics[width = 0.66\columnwidth]{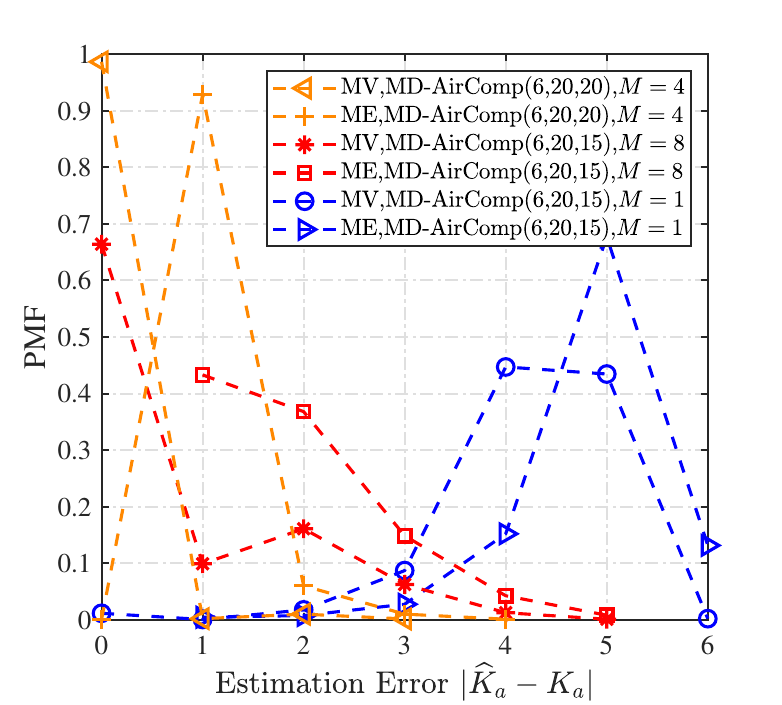}
\captionsetup{font={footnotesize, color = {black}}, singlelinecheck = off, justification = raggedright,name={Fig.},labelsep=period}
\caption{The PMF of the $K_a$ estimation error over 1000 training rounds.}  
\vspace{-8mm}
\label{FigKE}
\end{figure}

Fig. \ref{FigKE} illustrates the $K_a$ estimation error. We consider the probability mass function (PMF) of the majority voting (MV) method and the mean estimation (ME) method for different MD-AirComp schemes. Note that the PMF results are calculated based on one realization of the scheme and on 1000 samples (training rounds). It is obvious that the MV method estimates $K_a$ much better than the ME scheme. Furthermore, due to the worse NMSE performance, the $K_a$ estimation error of ``MD-AirComp (6, 20, 15), $M=1$" is much worse than that of ``MD-AirComp (6, 20, 15), $M=8$". Furthermore, Table \ref{TableSNR} compares the performance of ``MD-AirComp (6, 20, 20), $M=4$" scheme under various SNRs. It can be observed that the proposed MD-AirComp scheme still works well in low SNR, i.e., $\text{SNR}=0$\,dB or $5$\,dB, with a negligible loss in the test accuracy.

\begin{table}[t]\small
\centering
\captionsetup{font = {normalsize}, labelsep = period} 
\begin{threeparttable}[b]
\caption{Performance comparison under various SNRs}
\begin{tabular}{|c|c|c|c|}
\Xhline{1.2pt}
SNR & 0\,dB & 5\,dB & 20\,dB\\
\Xhline{1.2pt}
Test Accuracy ($t=1000$) & 0.8402 & 0.8495 & 0.8513\\
\hline
NMSE ($t=1000$) & -2.75\,dB & -5.33\,dB & -16.30\,dB\\
\hline
NMSE ($t=10$) & -3.67\,dB & -7.05\,dB & -22.7\,dB\\
\Xhline{1.2pt}
\end{tabular}
\label{TableSNR}
\end{threeparttable}
\end{table}

\begin{figure}[t] 
\centering  
\includegraphics[width = 0.66\columnwidth]{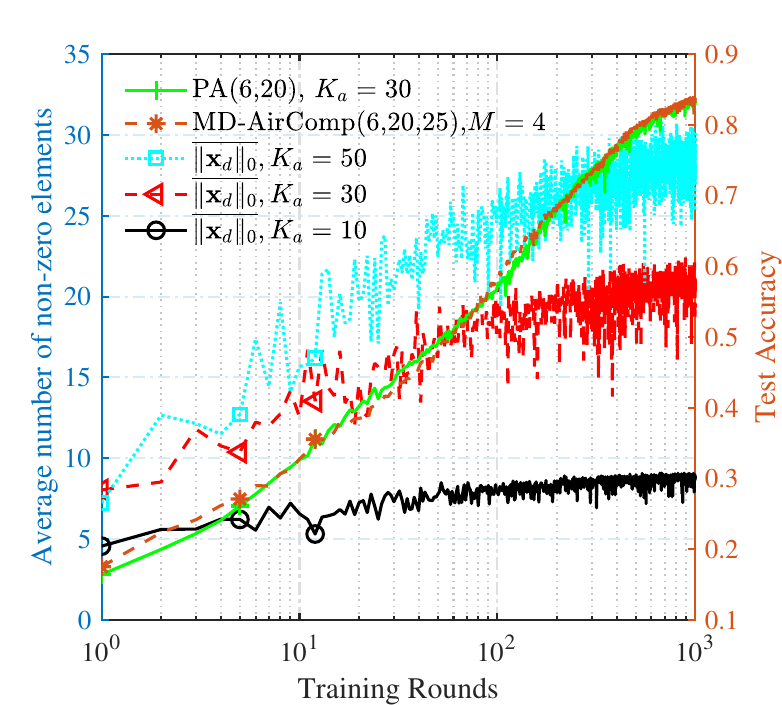}
\captionsetup{font={footnotesize}, singlelinecheck = off, justification = justified, name={Fig.},labelsep=period}
\caption{Performance evaluation when $K=100$. Left y-axis: Average number of non-zero elements $\|{\bf x}_d\|_0$ under different $K_a$; Right y-axis: Test accuracy of ``PA(6, 20)" and ``MD-AirComp(6, 20, 25)", where $K_a=30$.}  
\vspace{-3mm}
\label{FigNonZ}
\end{figure}

Fig. \ref{FigNonZ} investigates the performance of the proposed MD-AirComp in a larger cell with $K=100$ devices. This only changes the local training dataset setting, other parameters are the same unless otherwise specified. Specifically, each one of the 100 devices is randomly assigned 100 samples from the dataset, while the remaining 40,000 samples are sorted by their labels and grouped into 100 shards of size 400. We first evaluate the average number of non-zero elements, i.e., the average $\|{\bf x}_d\|_0$, denoted as $\overline{\|{\bf x}_d\|_0}\triangleq\frac{\sum_{d=1}^D\|{\bf x}_d\|_0}{D}$. We consider different device activity ratios, i.e., 10\%, 30\%, 50\%, and run the PA (6, 20) scheme to calculate the average $\|{\bf x}_d\|_0$. It can be seen that $\overline{\|{\bf x}_d\|_0}$ can be much smaller than $K_a$, indicating a high collision rate among active  devices (similar conclusion can be seen in Fig. \ref{FigWTa}). We reiterate that the collision helps to reduce the communication overhead of the proposed MD-AirComp scheme, and $L$ can be smaller than $K_a$ due to the collisions. As can be verified in Fig. \ref{FigNonZ}, considering $K_a=30$, ``MD-AirComp (6, 20, 25), $M=4$", i.e., $L=25<K_a$, approaches the PA(6, 20) benchmark very well.

\begin{figure}[t] 
\centering  
\includegraphics[width = 0.66\columnwidth]{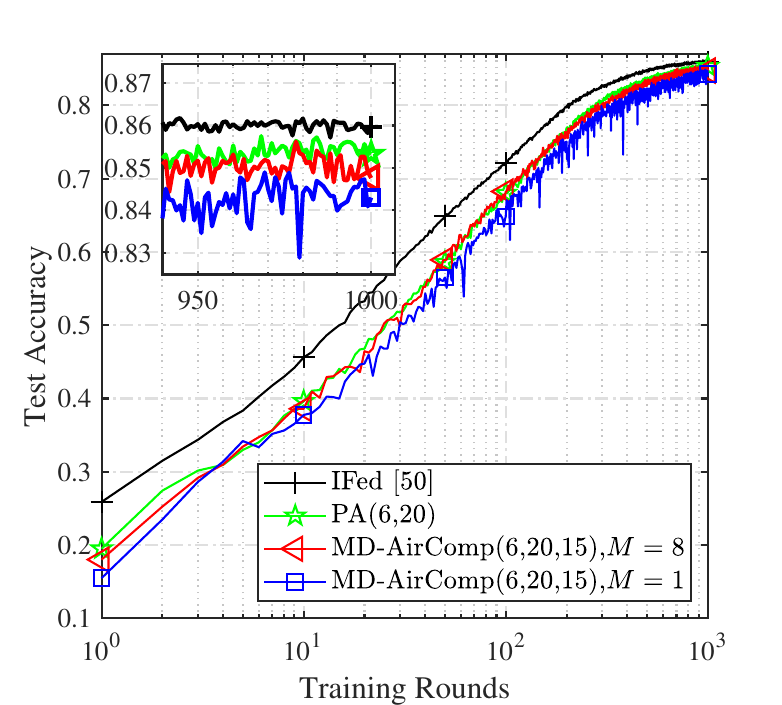}  
\captionsetup{font={footnotesize}, singlelinecheck = off, justification = raggedright,name={Fig.},labelsep=period}
\caption{Test accuracy versus the logarithmic training rounds.}  
\vspace{-4mm}
\label{FigTheory}
\end{figure}

Furthermore, in Fig. \ref{FigTheory}, we study effects of the various terms in the convergence analysis and Theorem \ref{thm:full-noncomp} through simulations. According to Corollary \ref{cor:fedams-full}, with the proper choice of the local and global learning rates, the convergence rate for MD-AirComp-based FEEL is $\cO\left(\frac{1}{\sqrt{T}}\right)$ and independent of the communication errors and quantization effects. This is confirmed in Fig. \ref{FigTheory} as we see a roughly similar slope in convergence of IFed, PA, and MD-AirComp schemes. Please note that, IFed and PA are communication error-free, i.e., the third term $\Psi$ in (\ref{eq:conve-thm}) is zero. IFed is also free of the quantization losses, i.e., $q=0$, indicating a smaller $C_1$ and a smaller second term $\Xi$ than that of PA. MD-AirComp has the third term that relate to both communication error $\sigma^2_n$ and quantization error $C_0$. From Fig. \ref{FigTheory}, the quantization effects seem to be more destructive. On the other hand, our proposed MD-AirComp technique seems to tackle the communication errors efficiently thereby achieving a performance very close to the PA benchmark.

\subsection{Robustness to Imperfections}
For simplicity, we focus on the received signal at the first antenna of the BS (pre-equalization applied at the device). 
\subsubsection{Imperfect Channel Estimation}

Here, we use ${\bf y}_{d}\in\mathbb{C}^{L}$ and ${\bf z}_{d}\in\mathbb{C}^{L}$ to denote the received signal and the noise of the $d$-th, $\forall d\in [D]$, block at the BS, respectively. In addition, denote ${h}_k$ and $\hat{h}_k$ as the actual channel and the estimated channel of the $k$-th device, respectively. Then, we can have the received signal model as
\begin{align}\label{ImpfCSI}
    {\bf y}_{d}&=\sum\nolimits_{k\in\mathcal{S}_a} \tilde{h}_k {\bf P}{\bf x}_k^d + {\bf z}_{d}\\ \nonumber
    &={\bf P}{\bf x}_{d} +\sum\nolimits_{k\in\mathcal{S}_a} (\tilde{h}_k-1) {\bf P}{\bf x}_k^d+ {\bf z}_{d},\\ \nonumber
    &={\bf P}{\bf x}_{d} + {\bf z}_{e} + {\bf z}_{d},
\end{align}
where $\tilde{h}_k=\frac{{h}_k}{\hat{h}_k}\in\mathbb{C}$ denotes the residual after pre-equalization, ${\bf z}_{e}$ is the error term caused by imperfect downlink channel estimation. If $\hat{h}_k={h}_k$, $\forall k\in\mathcal{S}_a$, ${\bf z}_{e}=0$ and (\ref{ImpfCSI}) simplifies to (\ref{TranMIMO}). According to (\ref{ImpfCSI}), even though there is channel estimation error, by seeing ${\bf z}_{e} + {\bf z}_{d}$ as a whole, we can still employ {\bf Algorithm 2} to estimate ${\bf x}_d$ and then aggregate the model. Note that we assume the noise obeys complex Gaussian distribution, and we estimate the mean and variance of such distribution iteratively by using EM algorithm. Similar to \cite{Yulin, ChenMZ,surveyAirComp}, we set the amplitude of $\tilde{h}_k$ as 1, $\forall k$, and focus on the phase error $\phi_k$. Specifically, we consider $\tilde{h}_k=e^{\phi_k}$ and $\phi_k$, $\forall k\in[K]$, is uniformly distributed in the set (0, $\phi$), where $\phi$ is the maximum phase offset. 

\subsubsection{Imperfect Time Synchronization}
To describe the effect of imperfect time synchronization, we follow the time-domain signal model of \cite{Yulin}. The received signal $r(\tilde{t})$ at the BS can be written as 
\begin{equation}\label{TimeSync}
r(\tilde{t}) = \sum\limits_{k\in\mathcal{S}_a} \sum_{l=1}^{L}  \left[{\bf P}{\bf x}_k^d \right]_l \text{rect}(\tilde{t} - \tilde{\tau}_k - l\tilde{T}) + z(\tilde{t}),
\end{equation}
where $\text{rect}(\tilde{t}) = \frac{1}{2}[\text{sgn}(\tilde{t} + \tilde{T}) - \text{sgn}(\tilde{t})]$ is a rectangular pulse of duration $\tilde{T}$, $\text{sgn}(\cdot)$ denotes the sign function, $\tilde{\tau}_k$ denotes the time offset of the $k$-th device, and $z(\tilde{t})$ is the noise. For simplicity, perfect channel gain pre-equalization is considered in (\ref{TimeSync}). Without loss of generality, we assume $\tilde{T}=1$ and the time offset $\tilde{\tau}_k$, $\forall k$, is less than the symbol duration $\tilde{T}$. We fix a maximum time offset $\tilde{\tau}$; then, we generate the time offset $\tilde{\tau}_k$, $\forall k$, uniformly in the set (0,$\tilde{\tau}$). Similar to \cite{Yulin,ChenMZ,surveyAirComp}, by applying whitened matched filtering and sampling scheme, we get oversampled, but independent, samples from the overlapped signals (\ref{TimeSync}). Then, the {\it aligned-sample estimator} (refer to Definition 3 of \cite{Yulin}) is applied in our simulation to estimate $\hat{\bf y}_d\triangleq \widehat{\sum\limits_{k\in\mathcal{S}_a}{\bf P}{\bf x}_k^d }$. Then, $\hat{\bf y}_d$ is used as input to conduct our proposed {\bf Algorithm 2}.

\begin{figure}[h]
 \vspace{-4mm}
\centering
\subfigure[]{
    \begin{minipage}[b]{1\linewidth}
        \centering
         \includegraphics[width=0.66\columnwidth]{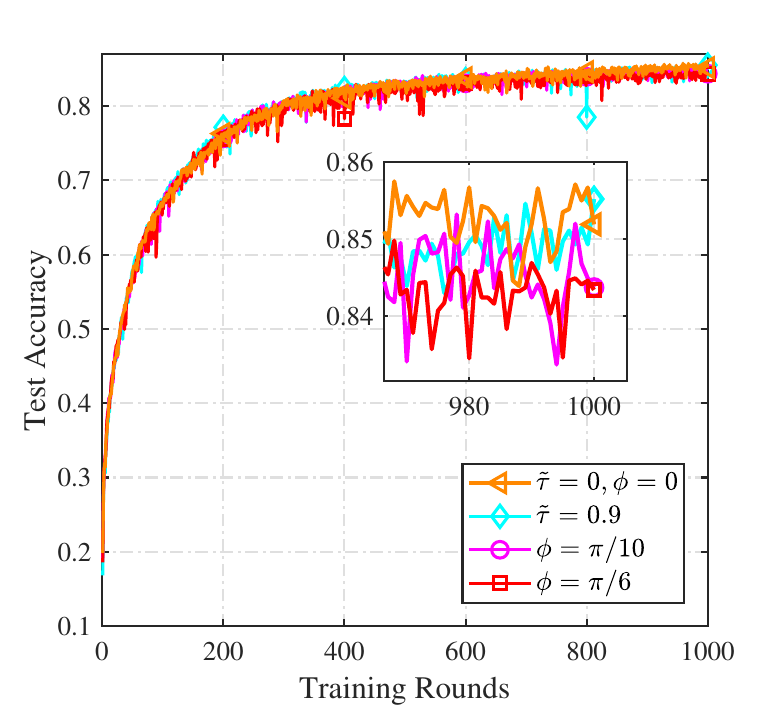}\\
\label{FigImpfa}
    \end{minipage}%
}\\
\vspace{-3.5mm}
\subfigure[]{
    \begin{minipage}[b]{1\linewidth}
        \centering
        \includegraphics[width=0.66\columnwidth]{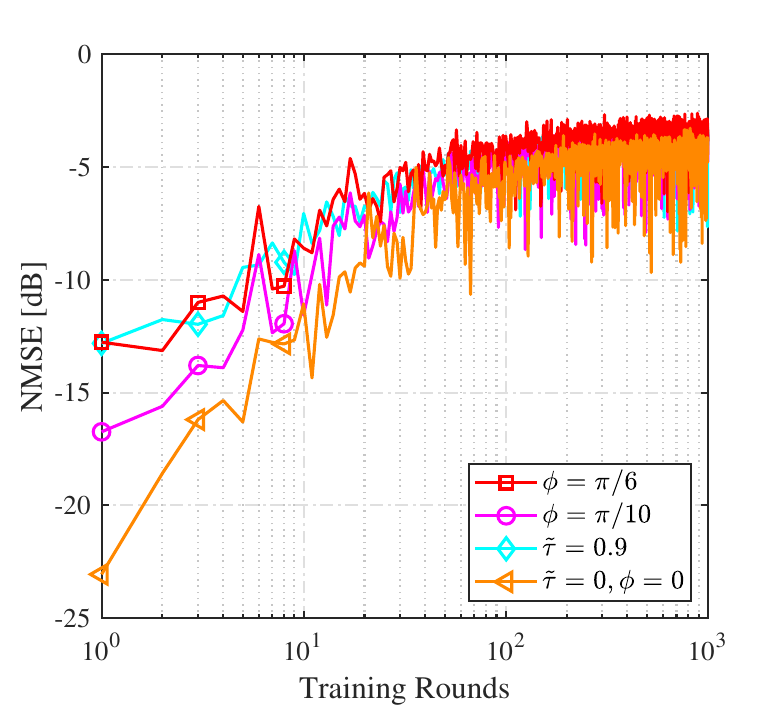}\\
\label{FigImpfb}
    \end{minipage}%
}%
\centering
\setlength{\abovecaptionskip}{-1mm}
\captionsetup{font={footnotesize}, singlelinecheck = off, justification = justified,name={Fig.},labelsep=period}
\caption{Performance of the proposed MD-AirComp scheme under various imperfections: (a) Test accuracy vs training rounds; (b) Detection NMSE vs training rounds.}
\label{FigImpf}
 \vspace{-6mm}
\end{figure}

In Fig. \ref{FigImpf}, we investigate the performance of the proposed ``MD-AirComp (6, 20, 20), $M=1$" scheme under various imperfections, where ``$\phi=\pi/6$", ``$\phi=\pi/10$", ``$\tilde{\tau}=0.9$", and ``$\tilde{\tau}=0,\phi=0$" are considered. From Fig. \ref{FigImpfa}, it can be seen that the performance still converges to the same level of test accuracy at negligible loss of convergence speed. From Fig. \ref{FigImpfb}, it is shown that both increasing the time offset and increasing the phase offset result in an SNR penalty, leading to degradation in NMSE performance of the proposed AMP-DA algorithm. Fortunately, the degraded estimation accuracy of AMP-DA algorithm is still acceptable for the FEEL task. In addition, as has been concluded from Fig. \ref{FigWTb}, it is also possible to increase $L$ and $M$ to further improve the NMSE performance of the AMP-DA algorithm. To sum up, our proposed MD-AirComp scheme is robust to imperfect channel estimation or imperfect time synchronization.

\section{Conclusions}
We have proposed a novel communication-efficient MD-AirComp scheme which can allow using over-the-air computation over legacy digital communication systems. The main idea of the proposed solution relies on the fact that, in AirComp, the receiver is only interested in the summation of the signals from multiple devices, but not the identities of the transmitting devices or their individual signals. The individual values to be summed are first quantized based on a common quantization codebook, then each quantized element is modulated into a transmit sequence selected from the common non-orthogonal modulation codebook. These transmit sequences from different devices overlap at the BS. Assuming that the number of transmitting devices is much less than the codebook size, the proposed sparsity promoting AMP-DA algorithm can efficiently compute the summation results by estimating which sequences have been transmitted, and by how many devices.

{Furthermore, we applied the proposed technique for efficient model aggregation in FEEL, which is one of the most exciting parallel and distributed optimization applications.} Here, the large-dimension model update is divided into blocks, and MD-AirComp is applied separately on each block. Our analysis proved a convergence rate of $\cO(\frac{1}{\sqrt{T}})$ for the proposed MD-AirComp-based FEEL scheme, where $T$ denotes the number of global training rounds. Simulation results verified that the proposed MD-AirComp scheme significantly outperforms the state-of-the-art schemes in test accuracy, and approaches the performance of ideal federated learning, despite using limited communication resources. Simulation results also verify the robustness of our proposed MD-AirComp scheme to imperfect channel estimation, imperfect time synchronization, and low SNR scenarios. There are various promising directions to explore, such as the development of adaptive quantization strategies (possibly layer-wise) for MD-AirComp, pursuing end-to-end optimization using deep learning, and investigating the applicability of MD-AirComp for privacy, among several other possibilities.


\appendix
\subsection{Proof of Proposition 1}
\label{proofP1}
For simplicity, we omit the subscriptions $n$ and $m$ in the derivations. By substituting (\ref{eq:BayesDetail1}), (\ref{eq:BayesDetail2}), and (\ref{eq:prior2}) into (\ref{eq:Bayes}), we obtain
\begin{align}\label{eq:ProofP1_0}
    &\!p\left(x| r\right) \\\nonumber
    &=\!\!\dfrac{1}{p\left(r\right)}{\cal CN}(r;x,\varphi)\left[(1\!-\!a)\delta\left( x\right)\!\!+\!a{\cal CN}\left( x; \!{\mu}_0,  \!{\tau}_0 \right)\right],\\\nonumber
    &=\!\!\dfrac{1}{p\left(r\right)}\bigg[ (1\!-\!a){\cal CN}(0;r,\varphi)\delta\left( x\right)+ \\ \nonumber
    &~~~~~~~~a {\cal CN}(0;r-{\mu}_0,\varphi+{\tau}_0){\cal CN}(x;\frac{{\mu}_0\varphi+r{\tau}_0}{\varphi+{\tau}_0},\frac{\varphi{\tau}_0}{\varphi+{\tau}_0}) \bigg].
\end{align}
Then, substituting $p(r)=(1\!-\!a){\cal CN}(0;r,\varphi)+ a{\cal CN}(x;r-{\mu}_0,\varphi+{\tau}_0)$ into (\ref{eq:ProofP1_0}), we can obtain
\begin{align}\label{eq:ProofP1_1}
    \!p\left(x| r\right)=(1\!-\!\pi)\delta\left( x\right)+ \pi {\cal CN}(x;\frac{{\mu}_0\varphi+r{\tau}_0}{\varphi+{\tau}_0},\frac{\varphi{\tau}_0}{\varphi+{\tau}_0}),
\end{align}
where $\pi=\dfrac{a}{a+(1-a)\frac{{\cal CN}(0;r,\varphi)}{{\cal CN}(0;r-{\mu}_0,\varphi+{\tau}_0)}}$. Hence, we can easily obtain (\ref{eq:post1-2})-(\ref{eq:post4-2}). Hence, proposition \ref{Proppp1} is proven.

\subsection{Proof of Theorem 1}
\label{proofT1}
According to (\ref{LossGra})-(\ref{GlobalWe}) and (\ref{GlobalWeEs}), we have ${\bf s}_k^t = \mathcal{C}(\Delta_k^t+{\bf e}_k^t)$, ${\bf e}_k^{t+1}=\Delta_k^t+{\bf e}_k^t-{\bf s}_k^t$, and $\widehat{\bf w}^{t+1}-\widehat{\bf w}^{t}= \eta \widehat{\bf s}^t$. To facilitate the proof, we introduce a new vector $\zb^{t+1} \triangleq \widehat{\wb}^{t +1} + \eta  \eb^{t+1}$, $\forall t\in[T]$, where $\eb^{t}=\frac{1}{K}\sum_{k=1}^K \eb_k^t$. We have $\zb^{t+1} - \zb^t = \eta(\Delta^{t} + \mb^t)$, where $\Delta^{t}\triangleq\frac{1}{K}\sum_{k=1}^K \Delta_k^t$ and the aggregation error is $\mb^t\triangleq{\widehat{\bf s}^{t}}-{\bf s}^{t}$. Note that the aggregation error $\mb^t$ results from the wireless transmission in FEEL.

According to Assumption \ref{as:smooth}, we have
\begin{align}\label{eq:f-Lsmooth}
& \EE[f(\zb^{t+1})]-f(\zb^{t}) \notag\\
& \leq \EE[\langle \nabla f(\zb^t), \zb^{t+1}-\zb^t \rangle] + \frac{\overline{L}}{2} \EE[\|\zb^{t+1}-\zb^t\|^2] \notag\\
& \leq \EE\left[\left\langle\nabla f(\zb^{t}), \eta   \Delta^{t}\right\rangle\right] +\frac{\eta^{2} \overline{L}}{2} \EE\left[\left\|  \Delta^{t}+ \mb^t\right\|^{2}\right] \notag\\
& = \underbrace{\EE\left[\left\langle\nabla f(\wb^{t}), \eta   \Delta^{t}\right\rangle\right]}_{R_1}  + \underbrace{\frac{\eta^{2} \overline{L}}{2} \EE\left[\left\|  \Delta^{t}+ \mb^t\right\|^{2}\right]}_{R_2}\notag\\
& +  \underbrace{\EE\left[\left\langle\nabla f(\zb^{t})-\nabla f(\wb^t), \eta  \Delta^{t}\right\rangle\right]}_{R_3}.
\end{align}

\textbf{Bounding $R_1$:}We have
\begin{align} \label{eq:T1-2}
    & R_1 = \EE\left[\left\langle\nabla f(\wb^{t}), \eta\Delta^{t}\right\rangle\right]\notag\\
    & = \eta\EE\left[\left\langle \nabla f(\wb^t) , \Delta^t + \eta_l {T_l} \nabla f(\wb^t) - \eta_l {T_l} \nabla f(\wb^t) \right\rangle\right] \notag\\
    & = -  \eta\eta_l {T_l}\EE\left[ \left\|\nabla f(\wb^t) \right\|^2\right] \!\!\!+\!  \eta\EE\left[\!\left\langle \!\nabla f(\wb^t) , \Delta^t \!\!+ \eta_l {T_l} \nabla f(\wb^t) \right\rangle\!\right] \notag\\
    & = - \eta\eta_l {T_l}\EE\left[ \left\|\nabla f(\wb^t) \right\|^2\right]  + \!\! \notag\\
&~~\underbrace{\eta \left\langle \nabla f(\wb^t),\EE \left[-\frac{\eta_l}{K}\sum_{k=1}^K \sum_{{t_l=0}}^{{T_l}-1} {\bf g}_{k, {t_l}}^t \!\!+\!\! \frac{\eta_l {T_l}}{K}\sum_{k=1}^K \nabla F_k(\wb^t)\right] \right\rangle}_{\widetilde{R_1}}.
\end{align}
Note that $f(\wb)= \frac{1}{K} \sum_{k=1}^K F_k(\wb)$, and ${\bf g}_{k, {t_l}}^t = \nabla F_k({\bf w}_{k, {t_l}}^t)$ is the local gradient. For the last term in \eqref{eq:T1-2}, by applying $\langle \xb,\yb\rangle = \frac{1}{2}(\|\xb\|^2 + \|\yb\|^2 - \|\xb-\yb\|^2)$ and the Cauchy-Schwarz inequality, we have 
\begin{align}\label{R1_p2}
    & \widetilde{R_1} =  \eta \bigg\langle \sqrt{\eta_l {T_l}} \nabla f(\wb^t),\notag\\
    &  - \frac{\sqrt{\eta_l {T_l}}}{{T_l} K} \EE \left[\sum_{k=1}^K \sum_{{t_l=0}}^{{T_l}-1} (\nabla F_k(\wb_{k, {t_l}}^t) - \nabla F_k(\wb^t))\right] \bigg\rangle \notag\\
    & \leq \frac{\eta \eta_l {T_l}}{2} \left\|\nabla f(\wb^t) \right\|^2  \!\!-\!\! \frac{\eta\eta_l}{2 {T_l} K^2} \EE \left[ \left\| \sum_{k=1}^K \sum_{{t_l=0}}^{{T_l}-1} \nabla F_i(\wb_{k, {t_l}}^t)\right\|^2\right]\notag\\
    & + \frac{\eta\eta_l}{2K}\sum_{k=1}^K \sum_{{t_l=0}}^{{T_l}-1}\EE \left[ \left\| \nabla F_i(\wb_{k, {t_l}}^t) - \nabla F_i(\wb^t) \right\|^2\right].
\end{align}
Then, by applying Lemma 3 of \cite{FedAdam} and using Assumption~\ref{as:smooth}, under the constraint of local learning rate $\eta_l \leq \frac{1}{8{T_l}\overline{L}}$, we have 
\begin{align} \label{R1_p3}
    & \frac{1}{K}\sum_{k=1}^{K}\EE \left[ \left\| \nabla F_i(\wb_{k, {t_l}}^t) - \nabla F_i(\wb^t) \right\|^2\right]\notag\\
    &\leq  \overline{L}^2\EE \left[\left\| \wb_{k, {t_l}}^t - \wb^t \right\|^2\right]  \notag\\
    & \leq 5{T_l}\eta_l^2\overline{L}^2(\sigma_l^2+6{T_l}\sigma_g^2)+ 30 {T_l}^2 \eta_l^2 \overline{L}^2\EE[\|\nabla f(\wb^t)\|^2].
\end{align}

Hence, by merging (\ref{eq:T1-2})-(\ref{R1_p3}) together, we obtain the bound
\begin{align}\label{eq:T1}
    R_1 & \leq -\frac{ \eta\eta_l {T_l}}{4} \EE\left[ \left\|\nabla f(\wb^t) \right\|^2\right] + \frac{5\eta\eta_l^3 {T_l}^2 \overline{L}^2}{2} (\sigma_l^2+6{T_l} \sigma_g^2)  \notag\\
    &~~~~~~~- \frac{ \eta\eta_l}{2 {T_l} K^2} \EE \left[ \left\|\sum_{k=1}^K \sum_{{t_l}=0}^{{T_l}-1} \nabla F_k(\wb_{k, {t_l}}^t)\right\|^2\right].
\end{align}

\textbf{Bounding $R_2$:} It can be bounded as follows:
\begin{align}\label{eq:T3}
    R_2 & = \frac{\eta^2 \overline{L} }{2} \EE\left[ \left\| \Delta^{t} + \mb^{t}\right\|^2\right] \notag\\
    & \leq \eta^2 \overline{L} \EE\left[ \left\|  \Delta^{t}\right\|^2\right]+ \eta^2 \overline{L}\EE\left[ \left\| \mb^{t} \right\|^2\right] ,
\end{align}
where the first inequality follows from Cauchy-Schwarz inequality. 
Referring to Section III-B, we can obtain 
\begin{align}\label{eq:T3_p2}
    &\EE\left[ \left\| \mb^{t} \right\|^2\right] \!=\! \EE\left[ \left\| \widehat{\bf s}^{t}\!-\!{\bf s}^{t} \right\|^2\right]\notag\\
    & =\EE\left[ \left\|\!\sum_{d=1}^{D}\left(\dfrac{1}{\widehat{K}^t} \Ub^t \widehat{\bf x}_d^t \!- \!\dfrac{1}{K^t} \Ub^t \xb_d^t \right)\right\|^2\right]\notag\\
    &\overset{(a)}{\leq} \dfrac{\| \Ub^t\|^2}{(K^t)^2}\sum_{d=1}^{D}\left\| \widehat{\bf x}_d^t-\xb_d^t\right\|^2 \overset{(b)}{=} \dfrac{\| \Ub^t\|^2 D N\sigma_n^2}{(K)^2},
\end{align}
where (a) and (b) hold by Assumption \ref{as:AMPasumption}. Note that (b) also assumes ${K}^t={K}$, i.e., the number of participating devices does not change along training rounds.

According to (\ref{Quant}), for a sufficiently large VQ codebook, the average power of $\Ub^t$ will be the average power of $\overline{\bf s}_k^t$, $\forall k\in[K]$. Hence, we have
\begin{align}\label{eq:T3_p2}
 \dfrac{\| \Ub^t\|^2 D N\sigma_n^2}{(K^t)^2} 
 \!\approx\!\dfrac{QN^2\| \overline{\bf s}_k^t\|^2 D \sigma_n^2}{W(K^t)^2} 
 \!\leq \!\dfrac{C_0\eta_l^2T_l^2G^2N^2\sigma_n^2}{(K^t)^2},
\end{align}
where the inequality holds by Assumptions \ref{as:bounded-g} and \ref{as:compressor}, and $C_0\triangleq\frac{2(1+q^2)^2}{(1-q^2)^2}$.

\textbf{Bounding $R_3$: }
\begin{align}
    R_3 & = \EE\left[\left\langle\nabla f(\zb^{t})-\nabla f(\wb^t), \eta   \Delta^{t}\right\rangle\right] \notag\\
    & \leq \EE\left[\|\nabla f(\zb^{t})-\nabla f(\wb^t)\|\left\|\eta   \Delta^{t}\right\|\right] \notag\\
    & \leq \overline{L} \EE\left[\|\zb^{t}-\wb^t\|\left\|\eta   \Delta^{t}\right\|\right] \notag\\
    & \leq \frac{\eta^2 \overline{L}}{2} \EE \left[\left\|  \Delta^{t}\right\|^2\right] + \frac{\eta^2 \overline{L}}{2} \EE\left[\left\|\eb^t \right\|^2 \right],
\end{align}
where the first inequality follows from the fact that $\langle \ab, \bb \rangle \leq \|\ab\|\|\bb\|$; the second one from Assumption \ref{as:smooth}; and the third from the definition of $\zb^t$ and the fact that $\|\ab\|\|\bb\| \leq \frac{1}{2}\|\ab\|^2 + \frac{1}{2}\|\bb\|^2$. Then, summing $R_3$ over $t=1, \ldots, T$, we have
\begin{align}\label{eq:T4-1}
    \sum_{t=1}^T R_3 & \leq \frac{\eta^2 \overline{L}}{2} \sum_{t=1}^T \EE \left[\left\|  \Delta^{t}\right\|^2\right] + \frac{\eta^2 \overline{L}}{2}\sum_{t=1}^T  \EE\left[\left\|  \eb^t \right\|^2 \right].
\end{align}
Then, by referring to Lemma C.7 of \cite{FedAMS}, the $R_3$ term is bounded by
\begin{align}\label{eq:T4-2}
    &\sum_{t=1}^T R_3  \leq \frac{\eta^2 \overline{L}}{2}  \sum_{t=1}^T \EE [\|\Delta_{t}\|^2]  + C_1 \eta^2 \overline{L} \sigma_l^2 \frac{T{T_l} \eta_l^2}{K} \notag \\
    & ~~~~+ C_1 \eta^2 \overline{L} \frac{\eta_l^2}{K^2} \sum_{t=1}^T \EE \left[\left\|\sum_{k=1}^K \sum_{{t_l}=0}^{{T_l}-1}\nabla F_i(\wb_{k, {t_l}}^t) \right\|^2\right],
\end{align}
where $C_1 = \frac{2(q+\gamma)^2}{(1-q^2)^2} $.

\textbf{Merging pieces together:}
Substituting \eqref{eq:T1} and \eqref{eq:T3} into \eqref{eq:f-Lsmooth}, summing over from $t=1$ to $T$, adding \eqref{eq:T4-2}, and then applying Lemma C.5 of \cite{FedAMS}, we have
\begin{align}
    & \EE[f(\zb^{T+1})]-f(\zb^1) = \sum_{t=1}^{T} [R_1+R_2+R_3] \notag\\
    & \leq -\frac{\eta\eta_l {T_l}}{4} \!\!\sum_{t=1}^{T} \EE\left[\left\|\nabla f(\wb^t) \right\|^2\right]\!+\!\! \frac{5\eta\eta_l^3 {T_l}^2 \overline{L}^2 T}{2} (\sigma_l^2+6{T_l} \sigma_g^2)  \notag\\
    &~~~~~~+ (\frac{3}{2}+C_1) \eta^2 \overline{L} \sigma_l^2 \frac{T{T_l} \eta_l^2}{K}+ \eta^2 \overline{L} \sum_{t=1}^{T} \dfrac{\| \Ub^t\|^2 DN \sigma_n^2}{(K)^2} +  \notag\\
    & \left(\!\frac{3\eta^2 \overline{L} \eta_l^2}{2K^2}\!\!-\! \!\frac{\eta\eta_l}{2 {T_l} K^2}\!\!+\!\!\frac{C_1\eta^2 \overline{L} \eta_l^2}{K^2}\!\right) \!\!\sum_{t=1}^{T} \!\EE\!\left[ \left\| \!\sum_{k=1}^K \!\!\sum_{{t_l}=0}^{{T_l}-1} \!\nabla \!F_i(\wb_{k, {t_l}}^t)\right\|^2\right]   \notag\\
    & \overset{(c)}{\leq} \!\!-\frac{\eta\eta_l {T_l}}{4} \!\!\sum_{t=1}^{T} \EE\left[\left\|\nabla f(\wb^t) \right\|^2\right]\! +\! \frac{5\eta\eta_l^3 {T_l}^2 \overline{L}^2 T}{2} (\sigma_l^2+6{T_l} \sigma_g^2) \notag\\
    & + \left(\!\frac{3}{2}+C_1\!\right) \eta^2 \overline{L} \sigma_l^2 \frac{T{T_l} \eta_l^2}{K} \!+ \! \dfrac{T\eta^2 \overline{L} C_0\eta_l^2T_l^2G^2N^2\sigma_n^2}{K^2},
\end{align}
where the inequality (c) holds since $\eta_l \leq \frac{1}{\eta L {T_l} (3+2C_1)}$.
Hence, we have 
\begin{align}\label{eq:fzT-3}
    & \frac{\eta\eta_l {T_l}}{4 T} \sum_{t=1}^{T} \EE[\|\nabla f(\wb^t)\|^2] \notag\\
    & \leq \frac{f(\zb^0)-\EE [f(\zb^T)]}{T} + \frac{5\eta\eta_l^3 {T_l}^2 \overline{L}^2 }{2} (\sigma_l^2+6{T_l} \sigma_g^2) \notag\\
    & + (\frac{3}{2}+C_1) \eta^2 \overline{L} \sigma_l^2 \frac{{T_l} \eta_l^2}{K} + \dfrac{\eta^2\eta_l^2 \overline{L}C_0T_l^2G^2N^2 \sigma_n^2}{K^2},
\end{align}
which implies 
\begin{align}
    \min_{t\in [T]} \EE [\|\nabla f(\wb^t)\|^2] \leq 4 \left[\frac{f_0-f_*}{\eta \eta_l {T_l} T} + \Xi + \Psi \right],
\end{align}
where $\Xi \triangleq (\frac{3}{2}+C_1) \eta \overline{L} \sigma_l^2 \frac{ \eta_l}{K} + \frac{5\eta_l^2 {T_l} \overline{L}^2 }{2} (\sigma_l^2+6{T_l} \sigma_g^2), \Psi \triangleq \frac{\eta\eta_l \overline{L}C_0T_lG^2N^2 \sigma_n^2}{K^2}$.

\end{document}